\documentclass[a4paper,12pt]{article}
\usepackage{amssymb, amsmath}
\usepackage[curve]{xypic}

\providecommand{\abs}[1]{\lvert#1\rvert}

\providecommand{\Spin}{\textnormal{Spin}}
\providecommand{\Pin}{\textnormal{Pin}}

\providecommand{\id}{\textnormal{id}}

\providecommand{\Ker}{\textnormal{Ker}}
\providecommand{\IIm}{\textnormal{Im}}
\providecommand{\Coker}{\textnormal{Coker}}

\providecommand{\Cyl}{\textnormal{Cyl}}
\providecommand{\Hom}{\textnormal{Hom}}

\providecommand{\Ker}{\textnormal{Ker}}

\providecommand{\Int}{\textnormal{Int}}

\providecommand{\SO}{\textnormal{SO}}
\providecommand{\OO}{\textnormal{O}}
\providecommand{\PP}{\textnormal{P}}
\providecommand{\Ab}{\textnormal{Ab}}

\begin{document}

\begin{titlepage}
\titlepage
\rightline{SISSA 47/2009/EP-FM}
\vskip 2.5cm
\centerline{ \bf \LARGE Revisiting pinors, spinors and orientability }
\vskip 1.7truecm

\begin{center}
{\bf \large Loriano Bonora, Fabio Ferrari Ruffino and Raffaele Savelli}
\vskip 1.5cm
\em 
International School for Advanced Studies (SISSA/ISAS) \\ 
Via Beirut 2, I-34151, Trieste, Italy\\
and Istituto Nazionale di Fisica Nucleare (INFN), sezione di Trieste

\vskip 2.5cm

\large \bf Abstract
\end{center}

\normalsize We study the relations between pin structures on a non-orientable even-dimensional manifold, with or without boundary, and pin structures on its orientable double cover, requiring the latter to be invariant under sheet-exchange. We show that there is not a simple bijection, but that the natural map induced by pull-back is neither injective nor surjective: we thus find the conditions to recover a full correspondence. We also show how to describe such a correspondence using spinors instead of pinors on the double cover: this is in a certain sense possible, but in a way that contains anyhow an explicit reference to pinors. We then consider the example of surfaces, with detailed computations for the real projective plane, the Klein bottle and the Moebius strip.

\vskip2cm

\vskip1.5\baselineskip

\vfill
 \hrule width 5.cm
\vskip 2.mm
{\small 
\noindent }
\begin{flushleft}
bonora@sissa.it, ferrariruffino@gmail.com, savelli@sissa.it
\end{flushleft}
\end{titlepage}

\newtheorem{Theorem}{Theorem}[section]
\newtheorem{Lemma}[Theorem]{Lemma}
\newtheorem{Corollary}[Theorem]{Corollary}
\newtheorem{Rmk}[Theorem]{Remark}
\newtheorem{Def}{Definition}[section]
\newtheorem{ThmDef}[Theorem]{Theorem - Defintion}

\tableofcontents

\section{Introduction}

Given a non-orientable manifold $X$ we call $\tilde{X}$ its orientable double cover, equipped with the orientation-reversing involution $\tau$ such that $\tilde{X} \,/\, \tau \simeq X$: we study the relations between pin structures on $X$ and $\tau$-invariant pin structures on $\tilde{X}$. We show that there is not a simple bijection as one might expect. In particular, recalling that there are two inequivalent euclidean pin groups at a fixed dimension, called $\Pin^{+}$ and $\Pin^{-}$, there is a natural map induced by pull-back:
	\[\Phi: \; \{\textnormal{pin$^{\pm}$ structures on } X \} \longrightarrow \{\textnormal{$\tau$-invariant } \textnormal{pin$^{\pm}$ structures on } \tilde{X} \}
\]
but this map is neither injective nor surjective. To make it surjective, we must impose one more condition: if $\tilde{\xi}$ is a pin$^{\pm}$ structure on $\tilde{X}$ and $\widetilde{d\tau}$ is an equivalence between $\tilde{\xi}$ and $\tau^{*}\tilde{\xi}$, then $\widetilde{d\tau}^{2}$ must be the identity, not the sheet-exchange of $\tilde{\xi}$ with respect to the tangent frame bundle $P_{O}\tilde{X}$. With this requirement we recover surjectivity. Moreover, non-injectivity is due to the fact that two pin$^{\pm}$ structures on $X$, which can be obtained from one another via the action of $w_{1}(X) \in H^{1}(X, \mathbb{Z}_{2})$, are pulled back to equivalent structures on $\tilde{X}$. This is clearer if we describe pin structures via the holonomy of the pin connection over 1-cycles: two such structures differ by the holonomy along the cycle whose lift in $\tilde{X}$ is not a cycle any more, but a path joining the two lifts of the same point of $X$. Thus, pulling them back to $\tilde{X}$ their difference disappears. We will see how to recover such a difference considering the correspondence, analogous to $\Phi$, for pinors as sections of the associated vector bundle, not simply for the pin structures themselves. Then, we will show how to describe the map $\Phi$ using spinors instead of pinors on $\tilde{X}$, exploiting the orientability of the latter: this is possible, but we will need anyhow to refer to pinors, in particular the difference between $\Pin^{+}$ and $\Pin^{-}$ will be preserved; this gives a global geometrical description of the approach considered in \cite{RvT}, and it is the explicit construction for pinors of what stated in \cite{BD} about a generic action of a discrete group on a manifold. In the second part of the paper we consider the analogous results for the case of manifolds with boundary.

The paper is organized as follows. In section 2 we briefly recall the fundamental notions about pin structures and we study the case of non-orientable manifolds without boundary. In section 3 we make explicit computations for surfaces, verifying directly what previously stated for the real projective plane and the Klein bottle. In section 4 we deal with manifolds with boundary, recalling the orientable case and then considering the non-orientable one, and we end with the explicit example of the Moebius strip.

\section{Pinors vs Spinors}

\subsection{Preliminaries on pinors}

We recall that the group $\SO(n)$ has a unique 2-covering $\Spin(n)$, while the group $\OO(n)$ has two inequivalent 2-coverings $\Pin^{\pm}(n)$, obtained from the Clifford algebras with positive and negative signature respectively, as explained in \cite{KT} (for Clifford algebras we use the convention $vw + wv = 2\langle v, w \rangle$, without the minus sign). Let $p^{\pm}: \Pin^{\pm}(n) \rightarrow \OO(n)$ be such 2-coverings with kernel $\{\pm 1\}$, both restricting to $\rho: \Spin(n) \rightarrow \SO(n)$. If we fix a the canonical basis $\{e_{1}, \ldots, e_{n}\}$ of $\mathbb{R}^{n}$ and we denote by $j_{1}$ the reflection with respect to the hyperplane $e_{1}^{\bot}$, we have that $\OO(n) = \langle \SO(n), j_{1} \rangle$, and $(p^{\pm})^{-1}(\{1,j_{1}\}) = \{\pm 1, \pm e_{1}\}$: the latter is isomorphic to $\mathbb{Z}_{4}$ if $e_{1}^{2} = -1$ and to $\mathbb{Z}_{2} \oplus \mathbb{Z}_{2}$ if $e_{1}^{2} = 1$, that's why in general we get non-isomorphic coverings. For details the reader can see \cite{KT}.
\begin{Def} Let $\pi: E \rightarrow M$ be a vector bundle of rank $n$ with a metric and let $\pi_{\OO}: \PP_{\OO}E \rightarrow M$ be the principal $\OO(n)$-bundle of orthonormal frames. A \emph{pin$^{\pm}$ structure} on $E$ is a principal $\Pin^{\pm}(n)$-bundle $\pi_{\Pin^{\pm}}: \PP_{\Pin^{\pm}}E \rightarrow M$ with a 2-covering $\xi: \PP_{\Pin^{\pm}}E \rightarrow \PP_{\OO}E$ such that the following diagram commutes:
	\[\xymatrix{
	\PP_{\Pin^{\pm}}E \times \Pin^{\pm}(n) \ar[r]^{\qquad\cdot} \ar[dd]_{\xi \times p^{\pm}} & \PP_{\Pin^{\pm}}E \ar[dd]^{\xi} \ar[dr]^{\pi_{\Pin^{\pm}}} \\
	& & M \; . \\
	\PP_{\OO}E \times \OO(n) \ar[r]^{\quad\;\cdot} & \PP_{\OO}E \ar[ur]_{\pi_{\OO}}
}\]
\end{Def}
\begin{Def} Two pin$^{\pm}$ structures $(\pi_{\Pin^{\pm}}, \xi^{\pm})$ and $(\pi'_{\Pin^{\pm}}, \xi'^{\pm})$ are \emph{equivalent} if there exists a principal $\Pin^{\pm}$-bundles isomorphism $\varphi: \PP_{\Pin^{\pm}}E \rightarrow \PP'_{\Pin^{\pm}}E$ such that $\xi' \circ \varphi = \xi$.
\end{Def}
Similarly to the case of spin structures, there is a simply transitive action of $H^{1}(M, \mathbb{Z}_{2})$ on pin$^{\pm}$ structures on a bundle $E \rightarrow M$. Given a good cover $\mathfrak{U} = \{U_{\alpha}\}_{\alpha \in I}$ of $M$, the bundle $E$ is represented by $\OO(n)$-valued transition functions $\{g_{\alpha\beta}\}$. A pin$^{\pm}$ structure is represented by $\Pin^{\pm}(n)$-valued transition functions\footnote{It is not true that the pin structure depends only on $[\{s_{\alpha\beta}\}\,] \in H^{1}(M, \underline{\Pin}^{\pm}(n))$ for $\underline{\Pin}^{\pm}(n)$ the sheaf of $\Pin^{\pm}$-valued smooth functions, because such a cohomology class determines the equivalence class of the principal bundle without considering the projection to the tangent bundle. In particular, two spin lifts can be isomorphic as principal pin$^{\pm}$-bundles, but in such a way that there are no isomorphisms commuting with the projections to $P_{O}M$: in this case they determine the same class in $H^{1}(M, \underline{\Pin}^{\pm}(n))$ but they are not equivalent as pin structures.} $\{s_{\alpha\beta}\}$ such that $p^{\pm}(s_{\alpha\beta}) = g_{\alpha\beta}$; all other pin$^{\pm}$ structures are represented by $\{s_{\alpha\beta} \cdot \varepsilon_{\alpha\beta}\}$ for $\{\varepsilon_{\alpha\beta}\}$ a $\mathbb{Z}_{2}$-cocycle and depend up to equivalence only by $[\{\varepsilon_{\alpha\beta}\}] \in \check{H}^{1}(M, \mathbb{Z}_{2})$. In particular, this implies that if there exist both $\Pin^{+}$ and $\Pin^{-}$ structures, their number is the same. Given a real vector bundle $\pi: E \rightarrow M$ the following conditions hold:
\begin{itemize}
	\item $E$ admits a $\Pin^{+}$-structure if and only if $w_{2}(E) = 0$;
	\item $E$ admits a $\Pin^{-}$-structure if and only if $w_{2}(E) + w_{1}(E) \cup w_{1}(E) = 0$.
\end{itemize}
For the proof the reader is referred to \cite{Karoubi}. As for spin structures, a pin structure on a manifold is by definition a pin structure on its tangent bundle.

\paragraph{}Let $M$ be a manifold of dimension $2n$. Given a pin$^{\pm}$ structure $\xi: P_{\Pin^{\pm}}M \rightarrow P_{\OO}M$ and an isometry $\varphi: M \rightarrow M$, we define the pin$^{\pm}$ structure $\varphi^{*}\xi$ via the following diagram:
	\[\xymatrix{
	& P_{\Pin^{\pm}}M \ar[d]^{\xi} \ar[dl]_{\varphi^{*}\xi} \\
	P_{\OO}M \ar[d]_{p} \ar[r]^{d\varphi} \ar[r]^{d\varphi} & P_{\OO}M \ar[d]^{p}  \ar@/^/[l] \\
	M \ar[r]^{\varphi} & M
}\]
i.e.\ $\varphi^{*}\xi = d\varphi^{-1} \circ \xi$. We remark that total space of the principal bundle $P_{\Pin^{\pm}}M$ is the same for both $\xi$ and $\varphi^{*}\xi$: what changes is the way it covers $P_{\OO}M$. Thus, also the vector bundle of pinors $S := P_{\Pin^{\pm}}M \times_{\rho} \mathbb{C}^{2^{n}}$, being $\rho$ the standard action of $\Pin^{\pm}(2n)$ on $\mathbb{C}^{2^{n}}$ as Clifford module, has the same total space in both cases. The only difference is the projection to $M$, which can be seen ignoring the second line of the previous diagram: we simply have $p_{\varphi^{*}\xi} = \varphi^{-1} \circ p_{\xi}$, i.e.\ if we call $q_{\xi}: S_{\xi} \rightarrow M$ and $q_{\varphi^{*}\xi}: S_{\varphi^{*}\xi} \rightarrow M$ the two bundle of pinors of $\xi$ and $\varphi^{*}\xi$, it follows that $S_{\xi} = S_{\varphi^{*}\xi}$ as total spaces and $(S_{\varphi^{*}\xi})_{x} = (S_{\xi})_{\varphi(x)}$, i.e.\ $S_{\varphi^{*}\xi} = \varphi^{*}S_{\xi}$. This is the well-known geometrical property that pinors, as well as spinors, are scalars under isometries (or in general under diffeomorphisms \cite{DP}). The differential $d\varphi$ does not have any local effect on pinors (contrary to vectors), it just determines the way they must be globally thought of as pinors, i.e.\ the way the corresponding principal bundle covers $P_{\OO}M$.

\paragraph{}We recall that two pin$^{\pm}$ structures $\xi: P_{\Pin^{\pm}}M \rightarrow P_{\OO}M$ and $\xi': P'_{\Pin^{\pm}}M \rightarrow P_{\OO}M$ are equivalent if there exists a principal bundle isomorphism $\rho: P_{\Pin^{\pm}}M \rightarrow P'_{\Pin^{\pm}}M$ such that $\xi = \xi' \circ \rho$. We say that a pin$^{\pm}$ structure $\xi$ is invariant under an isometry $\varphi$ if $\xi \simeq \varphi^{*}\xi$, i.e.\ if there exists a (non-canonical) lift $\widetilde{d\varphi}$ completing the following diagram:
\begin{equation}\label{EquivalencePhiStar}
	\xymatrix{
	P_{\Pin^{\pm}}M \ar[d]_{\xi} \ar@{.>}[r]^{\widetilde{d\varphi}} & P_{\Pin^{\pm}}M \ar[d]^{\xi} \ar[dl]_{\varphi^{*}\xi} \\
	P_{\OO}M \ar[d]_{p} \ar[r]^{d\varphi} \ar[r]^{d\varphi} & P_{\OO}M \ar[d]^{p} \\
	M \ar[r]^{\varphi} & M.
}
\end{equation}
If such a $\widetilde{d\varphi}$ exists, there are only two possibilites, linked by an exchange of the two sheets: in fact, $\widetilde{d\varphi}$ is a lifting of the map $\varphi^{*}\xi$ to a $2:1$ covering of the codomain (\cite{Hatcher} prop. 1.34 pag. 62). Calling $\gamma$ the sheet exchange, the two possible liftings are $\widetilde{d\varphi}$ and $\widetilde{d\varphi} \circ \gamma$. Then $\widetilde{d\varphi} \circ \gamma = \gamma \circ \widetilde{d\varphi}$ since, if $\widetilde{d\varphi}(p_{x}) = q_{\varphi(x)}$, then the only possibility is that $\widetilde{d\varphi}(\gamma(p_{x})) = \gamma(q_{\varphi(x)})$ in order to cover $d\varphi$. 

We now consider invariance under isometries of pinors, i.e.\ of sections of the associated vector bundle. Since we just have $S_{\varphi^{*}\xi} = \varphi^{*}S_{\xi}$ so that the total spaces are the same, also the sections of the two bundles, as subsets of their total spaces, are the same. In particular, a section $s \in \Gamma(S_{\xi}^{+})$ becomes naturally a section of $\varphi^{*}S_{\xi}$ via the natural map:
	\[\begin{split}
	\eta_{\varphi}:\; & \Gamma(S_{\xi}) \longrightarrow \Gamma(S_{\varphi^{*}\xi})\\
	& s \longrightarrow \eta_{\varphi}(s): \; \eta_{\varphi}(s)_{x} := s_{\varphi^{-1}(x)} \; .
\end{split}\]
This is the scalar behaviour of a pinor field. If $\xi \simeq \varphi^{*}\xi$, from diagram \eqref{EquivalencePhiStar} we have the isomorphism $\widetilde{d\varphi}: P_{\Spin}M \rightarrow P_{\Spin}M$, unique up to sheet exchange, so we have a vector bundle map $\widetilde{d\varphi}: S_{\varphi^{*}\xi} \rightarrow S_{\xi}$: the ambiguity of $\widetilde{d\varphi}$ corresponds via $\rho$ to a sign ambiguity on the action on $S_{\xi}$. Thus we have a map:
	\[S_{\xi} \overset{\id}\longrightarrow S_{\varphi^{*}\xi} \overset{\pm\widetilde{d\varphi}}\longrightarrow S_{\xi}
\]
whose behaviour with respect to the base point is:
	\[(S_{\xi})_{x} \overset{\id}\longrightarrow (S_{\xi})_{x} = (S_{\varphi^{*}\xi})_{\varphi^{-1}(x)} \overset{\pm\widetilde{d\varphi}}\longrightarrow (S_{\xi})_{x}
\]
(note that $\widetilde{d\varphi}$ commutes with projections since it is a bundle map, while the identity is not) inducing the natural map of sections:
	\[\xymatrix{
	\Gamma(S_{\xi}) \ar[r]^{\eta_{\varphi}} & \Gamma(S_{\varphi^{*}\xi}) \ar[r]^{\pm\widetilde{d\varphi}} & \Gamma(S_{\xi})
}\]
so that the invariance condition reads:
	\[s = \pm \widetilde{d\varphi} \circ \eta_{\varphi}(s)
\]
i.e.\ $s_{x} = \pm\widetilde{d\varphi}(s_{\varphi^{-1}(x)})$. We remain with a sign ambiguity, contrary to the case of vectors, in which case we can completely define invariance of a section by requiring that $d\varphi(v) = v$. As explained in \cite{DP}, for pinors (as well as for spinors) we have just a \emph{projective} action of $\varphi$. So also the notion of invariance is affected by this.

\subsection{Double covering of a non-orientable manifold}

As is well-known, every non-orientable manifold $X$ has an orientable double-cover $\tilde{X}$ with an orientation-reversing involution $\tau$ such that $X \simeq \tilde{X} \,/\, \tau$. It can be constructed as follows: we choose an atlas $\{(U_{\alpha}, \psi_{\alpha})\}_{\alpha \in I}$ of $X$ with corresponding transition functions $g_{\alpha\beta}$, and we consider the $\mathbb{Z}_{2}$-bundle with charts $U_{\alpha} \times \mathbb{Z}_{2}$ and transition functions $\varepsilon_{\alpha\beta}$ equal to the sign of the Jacobian $J(g_{\alpha\beta})$. The involution $\tau$ is the exchange of the two sheets. If we consider the projection $\pi: \tilde{X} \rightarrow X$, then $\tilde{X}$ as a manifold has an atlas given by couples of charts $\{(\pi^{-1}U_{\alpha}, \psi_{\alpha} \circ \pi)\}_{\alpha \in I}$ so that the transition functions are still $g_{\alpha\beta}$ for both the components of $\pi^{-1}(U_{\alpha\beta})$. We now want to study the behaviour of $\pi_{*}: H_{1}(\tilde{X}, \mathbb{Z}_{2}) \rightarrow H_{1}(X, \mathbb{Z}_{2})$ and consequentely of $\pi^{*}: H^{1}(X, \mathbb{Z}_{2}) \rightarrow H^{1}(\tilde{X}, \mathbb{Z}_{2})$. We recall the following canonical isomorphisms (see \cite{Hatcher}):
\begin{equation}\label{CanonicalIso}
\begin{split}
	&H_{1}(X, \mathbb{Z}) \simeq \Ab \, \pi_{1}(X)\\
	&H_{1}(X, \mathbb{Z}_{2}) \simeq H_{1}(X, \mathbb{Z}) \otimes_{\mathbb{Z}} \mathbb{Z}_{2}\\
	&H^{1}(X, \mathbb{Z}_{2}) \simeq \Hom(H_{1}(X, \mathbb{Z}), \mathbb{Z}_{2}) \simeq \Hom(H_{1}(X, \mathbb{Z}_{2}), \mathbb{Z}_{2}) \; .
\end{split}
\end{equation}
Let us show that, for $\pi_{*}: H_{1}(\tilde{X}, \mathbb{Z}_{2}) \rightarrow H_{1}(X, \mathbb{Z}_{2})$, the image $\IIm \, \pi_{*}$ has always index two in $H_{1}(X, \mathbb{Z}_{2})$, or equivalently that $\Ker \, \pi^{*} \simeq \mathbb{Z}_{2}$ in $H^{1}(X, \mathbb{Z}_{2})$. This can be seen in two ways. One is the following simple algebraic lemma:
\begin{Lemma} Let $G$ be a group and $H$ a subgroup such that $[G:H] = 2$. Then the natural map $\Ab\,H \rightarrow \Ab\,G$ has image of index $2$.
\end{Lemma}
\textbf{Proof:} For $g \in G$, $g$ and $g^{-1}$ lie in the same $H$-coset. Thus, considering a commutator $g_{1}g_{2}g_{1}^{-1}g_{2}^{-1}$ we have that an even number of factors can lie in the coset $G \setminus H$, thus the product lives in $H$; hence $G' \leq H$. Let us prove that the image of the natural map $\psi: H / H' \rightarrow G / G'$ has index $2$. If $g' = gh$ for $g, g' \in G$ and $h \in H$, then $[g']_{G/G'} = [g]_{G/G'}[h]_{G/G'} = [g]_{G/G'}\psi([h]_{H/H'})$. Thus the number of cosets of $\IIm \psi$ in $G/G'$ is less than $2$. Moreover, if $g \in G\setminus H$ it cannot happen that $[g]_{G/G'} = \psi([h]_{H/H'})$, because otherwise $g \in h \cdot G' \subset h \cdot H = H$. $\square$

\paragraph{}The previous lemma for $G = \pi_{1}(X)$ and $H = \pi_{1}(\tilde{X})$ implies that $[H_{1}(X, \mathbb{Z}) : \pi_{*}H_{1}(\tilde{X}, \mathbb{Z})] = 2$. The other way to prove the latter result is by means of the following exact sequence in cohomology, which can be found in \cite{MS}:
	\[\xymatrix{
	\cdots \ar[r] & H^{i-1}(X, \mathbb{Z}_{2}) \ar[r]^{\cup w_{1}(X)} & H^{i}(X, \mathbb{Z}_{2}) \ar[r]^{\pi^{*}} & H^{i}(\tilde{X}, \mathbb{Z}_{2}) \ar[r] & H^{i}(X, \mathbb{Z}_{2}) \ar[r] & \cdots \; .
}\]
For $i = 1$, since $H^{0}(X, \mathbb{Z}_{2}) = \mathbb{Z}_{2}$ we have that $\IIm(\cup w_{1}(X)) = w_{1}(X)$, thus by exactness $\Ker \, \pi^{*} = \{0, w_{1}(X)\} \simeq \mathbb{Z}_{2}$. This is what we expected: since the double covering is orientable, the pull-back $\pi^{*}$ must kill $w_{1}(X)$.

\paragraph{}Since $\mathbb{Z}_{2}$ is a field, $H^{1}(\tilde{X}, \mathbb{Z}_{2})$ is a vector space, thus every subspace can be complemented. Hence we have:
\begin{equation}\label{SplitH1Z2}
	H^{1}(\tilde{X}, \mathbb{Z}_{2}) = \mathbb{Z}_{2}^{k} \oplus \IIm \, \pi^{*}
\end{equation}
where $\pi^{*}: H^{1}(X, \mathbb{Z}_{2}) \rightarrow \IIm \, \pi^{*}$ is a surjection with kernel $\mathbb{Z}_{2}$. Since $H^{1}(X, \mathbb{Z}_{2}) = H^{1}(X, \mathbb{Z}) \otimes_{\mathbb{Z}} \mathbb{Z}_{2}$, its dimension is equal to the Betti number $b_{1}(X)$ plus the number of $\mathbb{Z}_{2^{k}}$-components of $H^{1}(X, \mathbb{Z})$: we call this number $b_{1}^{(2)}(X)$. The dimension of $\IIm \, \pi^{*}$ is thus $b_{1}^{(2)}(X)-1$ so, in \eqref{SplitH1Z2}, we have $k = b_{1}^{(2)}(\tilde{X})-b_{1}^{(2)}(X)+1$. Thus we get the following general picture:
	\[\xymatrix{
	\mathbb{Z}_{2}^{\oplus ( b_{1}^{(2)}(X)-1 )} \oplus (\mathbb{Z}_{2}^{\oplus ( b_{1}^{(2)}(\tilde{X})-b_{1}^{(2)}(X)+1 )} \simeq \Coker\, \pi^{*}) \; \,\\
	\mathbb{Z}_{2}^{\oplus (b_{1}^{(2)}(X)-1)} \oplus (\mathbb{Z}_{2} = \Ker\, \pi^{*}) \ar@<18ex>[u]^{\pi^{*}}_{\simeq} \;.\qquad\qquad\qquad\qquad
}\]

\paragraph{}In the sequel we will also need another general result: we compare the tangent bundle of $\tilde{X}$ and the tangent bundle of $X$. We recall that if $f: X \rightarrow Y$ is a continuous map and $p: E \rightarrow Y$ a fiber bundle, the pull-back $\pi_{2}: f^{*}E \rightarrow X$ is defined as the fiber product $E \times_{Y} X$ via $\pi$ and $f$, thus its elements are of the form $(e, x)$ with $p(e) = f(x)$. The projection is $\pi_{2}(e,x) = x$.
\begin{Lemma}\label{TangentLemma} For $\pi: \tilde{X} \rightarrow X$ the projection and $p: TX \rightarrow X$, $\tilde{p}: T\tilde{X} \rightarrow \tilde{X}$ the tangent bundles, there is the canonical bundle isomorphism:
	\[\begin{split}
	\varphi: \; &T\tilde{X} \overset{\simeq}\longrightarrow \pi^{*}TX\\
	&\varphi(v) = (d\pi(v), \tilde{p}(v)) \; .
\end{split}\]
Similarly for the orthogonal frame bundles, with respect to a metric $g$ on $X$ and its pull-back $\pi^{*}g$ on $\tilde{X}$, there is the canonical isomorphism:
	\[\begin{split}
	\varphi_{O}: \; &P_{O}\tilde{X} \overset{\simeq}\longrightarrow \pi^{*}P_{O}X\\
	&\varphi_{O}(x) = (d\pi(x), \tilde{p}(x)) \; .
\end{split}\]
\end{Lemma}
\textbf{Proof:} It is easy to verify that $\varphi$ is a well-defined bundle map. It also follows from the definition of pull-back:
\begin{equation}\label{PiCartesian}
	\xymatrix{
	T\tilde{X} \ar[d]_{\tilde{p}} \ar@{.>}[r]_{\varphi} \ar@/^1pc/[rr]^{d\pi} & \pi^{*}TX \ar[r]_{\pi_{1}} \ar[d]_{\pi_{2}} & TX \ar[d]^{p}\\
	\tilde{X} \ar@/_1pc/[rr]_{\pi} \ar[r]^{\id} & \tilde{X} \ar[r]^{\pi} & X
}
\end{equation}
By construction the map $\varphi$ lifts the identity of $\tilde{X}$ and it must satisfy $\pi_{1}\varphi(v) = d\pi(v)$, thus $\varphi(v) = (d\pi(v), \tilde{p}(v))$. We have to verify that it is an isomorphism on each fiber, i.e.\ that $d\pi_{x}$ is an isomorphism for each $x$: this is true since $\pi$ is a local diffeomorphism. The same considerations apply for frame bundles. $\square$

\subsection{Pinors on the double covering}\label{PinorsDouble}

We now want to compare pinors on a non-orientable manifold $X$ and pinors on its double covering $\tilde{X}$ which are $\tau$-invariant. We start with the following simple lemma:
\begin{Lemma} If $X$ admits a $\Pin^{+}$-structure \emph{or} a $\Pin^{-}$-structure then $\tilde{X}$ is spin.
\end{Lemma}
\textbf{Proof:} by lemma \ref{TangentLemma} we have that $w_{2}(\tilde{X}) = \pi^{*}w_{2}(X)$. Since $w_{1}(X) \in \Ker\,\pi^{*}$, we obtain $\pi^{*}w_{2}(X) = \pi^{*}(w_{2}(X) + w_{1}(X) \cup w_{1}(X))$, thus if there is a pin$^{\pm}$ structure we get $w_{2}(\tilde{X}) = 0$. $\square$

\paragraph{}Let us suppose that a pin$^{\pm}$ structure on $\tilde{X}$ is $\tau$-invariant. Thus we have two possible liftings of $d\tau$ in diagram \eqref{EquivalencePhiStar}:
	\[\xymatrix{
	P_{\Pin^{\pm}}\tilde{X} \ar[d]_{\tilde{\xi}} \ar[rr]^{\widetilde{d\tau}, \;\widetilde{d\tau} \circ \gamma} & & P_{\Pin^{\pm}}\tilde{X} \ar[d]^{\tilde{\xi}} \ar[dll]_{\tau^{*}\tilde{\xi}} \\
	P_{O}\tilde{X} \ar[d]_{\tilde{p}} \ar[rr]^{d\tau} & & P_{O}\tilde{X} \ar[d]^{\tilde{p}} \\
	\tilde{X} \ar[rr]^{\tau} & & \,\tilde{X}.
}\]
Since $\widetilde{d\tau} \circ \gamma = \gamma \circ \widetilde{d\tau}$, it follows that $(\widetilde{d\tau} \circ \gamma)^{2} = \widetilde{d\tau}^{2}$, and the latter can be only $\id$ or $\gamma$, since it is an auto-equivalence of $\tilde{\xi}$ which covers $d\tau^{2} = \id$.

\paragraph{} We would like to show that the pull-back of a pin$^{\pm}$ structure on $X$ is a pin$^{\pm}$ structure on $\tilde{X}$ which is $\tau$-invariant and such that $\widetilde{d\tau}^{2} = \id$. Then, $\tilde{X}$ being orientable, we will be able to reduce the structure group to $\Spin$. Let us consider the following diagram:
\begin{equation}\label{PinPullback}
	\xymatrix{
	& \pi^{*}P_{\Pin^{\pm}}X \ar[r]^{\pi_{1}} \ar[d]|{(\xi,\id)} \ar[dl]_{\tilde{\xi}} & P_{\Pin^{\pm}}X \ar[d]^{\xi} \\
	P_{O}\tilde{X} \ar[d]_{\tilde{p}} \ar@{.>}[r]_{\varphi} \ar@/^2pc/[rr]_{\qquad\qquad\qquad\qquad\; d\pi} & \pi^{*}P_{O}X \ar[r]_{\pi_{1}} \ar[d]_{\pi_{2}} \ar@/_/@{.>}[l] & P_{O}X \ar[d]^{p}\\
	\tilde{X} \ar@/_2pc/[rr]^{\pi} \ar[r]^{\id} & \tilde{X} \ar[r]^{\pi} & X
}
\end{equation}
where $\tilde{\xi}$ defines the pull-back on $\tilde{X}$ of the spin structure $\xi$ of $X$, for $\varphi$ defined in lemma \ref{TangentLemma}. Thus we consider as total space of the bundle exactly $\pi^{*}P_{\Pin^{\pm}}X$. We now see that $\tilde{\xi}$ is $\tau$-invariant. We recall that $\tau^{*}\tilde{\xi}$ is defined by:
	\[\xymatrix{
	& \pi^{*}\Pin^{\pm}X \ar[d]_{\tilde{\xi}} \ar[dl]_{\tau^{*}\tilde{\xi}} \\
	P_{O}\tilde{X} \ar[d]_{\tilde{p}} \ar[r]^{d\tau} & P_{O}\tilde{X} \ar[d]^{\tilde{p}} \\
	\tilde{X} \ar[r]^{\tau} & \tilde{X}
}\]
and we claim that $\tau^{*}\tilde{\xi} \simeq \tilde{\xi}$ via the two possible equivalences:
	\[\xymatrix{
	\pi^{*}P_{\Pin^{\pm}}X \ar[d]_{\tilde{\xi}} \ar[rr]^{(1,\tau), (\gamma,\tau)} & & \pi^{*}P_{\Pin^{\pm}}X \ar[d]^{\tilde{\xi}} \ar[dll]_{\tau^{*}\xi} \\
	P_{O}\tilde{X} \ar[d]_{\tilde{p}} \ar[rr]^{d\tau} & & P_{O}\tilde{X} \ar[d]^{\tilde{p}} \\
	\tilde{X} \ar[rr]^{\tau} & & \,\tilde{X}
}\]
where $\gamma$ is the exchange of sheets of $P_{\Pin^{\pm}}X$ with respect to $P_{O}X$, while $\tau$ is the exchange of sheets of $\tilde{X}$ with respect to $X$. In fact, by diagram \eqref{PinPullback} we have $\tilde{\xi}(p', \tilde{x}) = \varphi^{-1}\circ (\xi,\id)(p', \tilde{x}) = \varphi^{-1}(p, \tilde{x}) = d\pi^{-1}_{\tilde{x}}(p)$ where $\pi_{\tilde{x}}$ is $\pi$ restricted to a neighborhood of $\tilde{x}$ on which it is a diffeomorphism. Therefore, for $\varepsilon = 1, \gamma$:
	\[\begin{split}
	&d\tau \circ \tilde{\xi}(p', \tilde{x}) = d\tau(d\pi^{-1}_{\tilde{x}}(p)) = d(\tau \circ \pi^{-1}_{\tilde{x}})(p) = d(\pi^{-1}_{\tau(\tilde{x})})(p)\\
	&\tilde{\xi} \circ (\varepsilon, \tau)(p', \tilde{x}) = \tilde{\xi}(\varepsilon(p'), \tau(\tilde{x})) = d(\pi^{-1}_{\tau(\tilde{x})})(p)
\end{split}\]
so that the diagram commutes. In particular, we see that the two possible isomorphisms $\widetilde{d\tau} = (1, \tau), (\gamma, \tau)$ have the property that $\widetilde{d\tau}^{2} = 1$. We have thus constructed a function:
	\[\Phi: \; \{\textnormal{pin$^{\pm}$ structures on } X \} \longrightarrow \{\textnormal{pin$^{\pm}$ structures on } \tilde{X} \textnormal{ $\tau$-invariant with } \widetilde{d\tau}^{2} = 1\} \; .
\]

\paragraph{}We now show that $\Phi$ is surjective, i.e.\ that a $\tau$-invariant pin$^{\pm}$ structure $\tilde{\xi}$ on $\tilde{X}$ satisfying $\widetilde{d\tau}^{2} = 1$ is the pull-back of a pin$^{\pm}$ structure on $X$. The latter is:
	\[\xi: P_{\Pin^{\pm}}\tilde{X} \, / \, \widetilde{d\tau} \longrightarrow P_{O}\tilde{X} \, / \, d\tau \simeq P_{O}X \; .
\]
In more detail:
	\[\xymatrix{
		P_{\Pin^{\pm}}\tilde{X} \,/\, \widetilde{d\tau} \ar[d]_{[\tilde{\xi}]} \ar[dr]^{\xi} \\
		P_{O}\tilde{X} \,/\, d\tau \ar[d]_{[\pi_{2}]} \ar[r]_{\quad \simeq}^{\quad \nu} & P_{O}X \ar[d]^{p} \\
		\tilde{X} \,/\, \tau \ar[r]^{[\pi]}_{\simeq} & \; X
	}
\]
where $\nu([p]) = d\pi(p)$. From $\widetilde{d\tau}^{2} = 1$ we get that the quotient is a 2-covering of $P_{O}X$, otherwise we would obtain a 1-covering, i.e.\ a bundle isomorphism, since $\gamma = \widetilde{d\tau}^{2}$ would identify also the two points of the same fiber. To see that $\tilde{\xi} \simeq \pi^{*}\xi$, we use the equivalence:
\begin{equation}\label{Mu}
\xymatrix{
	P_{\Pin^{\pm}}\tilde{X} \ar[dr]_{\tilde{\xi}} \ar[rr]^{\mu \quad}_{\simeq \quad} & & \pi^{*}(P_{\Pin^{\pm}}\tilde{X} \,/\, \widetilde{d\tau}) \ar[dl]^{\varphi^{-1} \circ (\xi, \id)} \\
	& P_{O}\tilde{X}
}
\end{equation}
for $\mu(\tilde{p}_{\tilde{x}}) = ([\tilde{p}_{\tilde{x}}], \tilde{x})$. The inverse of $\mu$ is given by $\mu^{-1}([\tilde{p}_{\tilde{x}}], \tilde{x}) = \tilde{p}_{\tilde{x}}$ or equivalently $\mu^{-1}([\tilde{p}_{\tilde{x}}], \tau(\tilde{x})) = \widetilde{d\tau}(\tilde{p}_{\tilde{x}})$. The diagram is commutative: $\varphi^{-1}\circ (\xi, \id) \circ \mu(\tilde{p}_{\tilde{x}}) = \varphi^{-1}\circ (\xi, \id)([\tilde{p}_{\tilde{x}}], \tilde{x}) = \varphi^{-1}((\nu\circ[\xi])([\tilde{p}_{\tilde{x}}]), \tilde{x}) = \varphi^{-1}(d\pi(\tilde{\xi}(\tilde{p}_{\tilde{x}})), \tilde{x}) = \tilde{\xi}(\tilde{p}_{\tilde{x}})$.

\paragraph{}It is easy to show that $\Phi$ commutes via $\pi^{*}$ with the actions of $H^{1}(X, \mathbb{Z}_{2})$ and $H^{1}(\tilde{X}, \mathbb{Z}_{2})$. In fact, for $\xi: P_{\Pin^{\pm}}X \rightarrow P_{O}X$ a pin$^{\pm}$ structure, up to isomorphism we can view $\Phi(\xi)$ as $\pi^{*}\xi: \pi^{*}P_{\Pin^{\pm}}X \rightarrow \pi^{*}P_{O}X$. We fix a $\rm\check{C}$ech class $[\omega] \in \check{H}^{1}(\mathfrak{U}, \mathbb{Z}_{2})$ for a good cover $\mathfrak{U} = \{U_{\alpha}\}_{\alpha \in I}$ of $X$. If the transition function of $P_{\Pin^{\pm}}X$ are $s_{\alpha\beta}$ and we fix a representative $\omega$, then the new transition functions are $s_{\alpha\beta} \cdot \omega_{\alpha\beta}$. On the two components of $\pi^{-1}U_{\alpha\beta}$, the transition functions were both $s_{\alpha\beta}$ and they become both $s_{\alpha\beta} \cdot \omega_{\alpha\beta}$, i.e., $\omega$ acts on the transition functions of $\pi^{*}\xi$ exactly as $\pi^{*}\omega$. Since $[\pi^{*}\omega] = \pi^{*}[\omega]$, we get the claim. Thus we have a diagram:
	\[\begin{array}{ccc}
	\{\Pin^{\pm}\textnormal{-structures on } X \} & \overset{\Phi}\longrightarrow & \{\Pin^{\pm}\textnormal{-structures on } \tilde{X} \textnormal{ $\tau$-invariant with } \widetilde{d\tau}^{2} = 1\} \\
	\circlearrowleft & & \circlearrowleft \\
	\check{H}^{1}(X, \mathbb{Z}_{2}) & \overset{\pi^{*}}\longrightarrow & \check{H}^{1}(\tilde{X}, \mathbb{Z}_{2}).
\end{array}\]
This implies in particular that $\Phi^{-1}(\tilde{\xi})$ is made by two inequivalent pin$^{\pm}$ structures, obtainable from each other via the action of $w_{1}(X) \in \Ker\,\pi^{*}$. We will now show that the two inequivalent counterimages can be recovered as $P_{\Pin^{\pm}}\tilde{X} \, / \, \widetilde{d\tau}$ and $P_{\Pin^{\pm}}\tilde{X} \, / \, (\widetilde{d\tau} \circ \gamma)$, by proving that these two quotients are inequivalent. In fact, let us suppose that there exists an equivalence:
	\[\xymatrix{
	P_{\Pin^{\pm}}\tilde{X} \, / \, \widetilde{d\tau} \ar[rr]^{\rho} \ar[dr]_{\xi} & & P_{\Pin^{\pm}}\tilde{X} \, / \, (\widetilde{d\tau} \circ \gamma) \ar[dl]^{\xi'} \\
	& P_{O}
}\]
then it lifts to an equivalence of the pull-backs:
	\[\xymatrix{
	& P_{O}\tilde{X} \\
	\pi^{*}(P_{\Pin^{\pm}}\tilde{X} \, / \, \widetilde{d\tau}) \ar[rr]^{\tilde{\rho}} \ar[d]_{\pi_{1}} \ar[ur]^{\tilde{\xi}} & & \pi^{*}(P_{\Pin^{\pm}}\tilde{X} \, / \, (\widetilde{d\tau} \circ \gamma)) \ar[d]^{\pi_{1}} \ar[ul]_{\tilde{\xi}'} \\
	P_{\Pin^{\pm}}\tilde{X} \, / \, \widetilde{d\tau} \ar[rr]^{\rho} \ar[dr]_{\xi} & & P_{\Pin^{\pm}}\tilde{X} \, / \, (\widetilde{d\tau} \circ \gamma) \ar[dl]^{\xi'} \\
	& P_{O}X
}\]
but, being both the pull-backs equivalent to $P_{\Pin^{\pm}}\tilde{X}$ via \eqref{Mu}, the only two possibilities for $\tilde{\rho}$ are the following:
	\[\xymatrix{
	P_{\Pin^{\pm}}\tilde{X} \ar[r]^{\id, \gamma} \ar[d]_{\mu} & P_{\Pin^{\pm}}\tilde{X} \ar[d]^{\mu'} \\
	\pi^{*}(P_{\Pin^{\pm}}\tilde{X} \, / \, \widetilde{d\tau}) \ar[r]^{\tilde{\rho} \quad} \ar@/_/[u] & \pi^{*}(P_{\Pin^{\pm}}\tilde{X} \, / \, (\widetilde{d\tau} \circ \gamma)). \ar@/^/[u]
}\]
Let us show that none of the two can be a lift of $\rho$. In fact, if it were so, they would be of the form:
\begin{equation}\label{RhoLift}
\tilde{\rho}([p], \tilde{x}) = (\rho[p], \tilde{x})
\end{equation}
while:
	\[\xymatrix{
	([p_{\tilde{x}}], \tilde{x}) \ar[r]^{\quad \mu^{-1}} & p_{\tilde{x}} \ar[r]^{\id} & p_{\tilde{x}} \ar[r]^{\mu' \quad} & ([p_{\tilde{x}}], \tilde{x}) \\
	([p_{\tilde{x}}], \tau(\tilde{x})) \ar[r]^{\quad \mu^{-1}} & \widetilde{d\tau}(p_{\tilde{x}}) \ar[r]^{\id} & \widetilde{d\tau}(p_{\tilde{x}}) \ar[r]^{\mu' \quad} & ([\widetilde{d\tau}(p_{\tilde{x}})], \tau(\tilde{x}))
}\]
and in the codomain $[p_{\tilde{x}}] \neq [\widetilde{d\tau}(p_{\tilde{x}})]$ since the class is taken with respect to $\widetilde{d\tau} \circ \gamma$, thus \eqref{RhoLift} is inconsistent. The same would happen choosing $\gamma$ instead of the identity. Thus $\tilde{\rho}$ lifts only the autoequivalences of each of the two quotients, not an equivalence between them.

\paragraph{}Now that we have seen the relationship between pin$^{\pm}$ structures on $X$ and the corresponding ones on $\tilde{X}$, we analyze such a relationship at the level of pinors (i.e.\ sections of the associated vector bundles). Let us start from $X$ and pull-back a pin$^{\pm}$ structure as in the following diagram:
	\[\xymatrix{
	\pi^{*}P_{\Pin^{\pm}}X \ar[d]_{\tilde{\xi}} \ar[r]^{\;\; \pi_{1}} & P_{\Pin^{\pm}}X \ar[d]^{\xi} \\
	P_{O}\tilde{X} \ar[r]^{d\pi} & P_{O}X.
}\]
For the associated bundles of pinors, we have that $(\pi^{*}P_{\Pin^{\pm}}X) \times_{\rho} \mathbb{C}^{2^{n}} \simeq \pi^{*}(P_{\Pin^{\pm}}X \times_{\rho} \mathbb{C}^{2^{n}})$ canonically. Thus, given on $X$ a pinor $s \in \Gamma(P_{\Pin^{\pm}}X \times_{\rho} \mathbb{C}^{2^{n}})$, we can naturally consider on $\tilde{X}$ its pull-back $\pi^{*}s \in \Gamma((\pi^{*}P_{\Pin^{\pm}}X) \times_{\rho} \mathbb{C}^{2^{n}})$. The natural equivalence between $\tilde{\xi}$ and $\tau^{*}\tilde{\xi}$ is given by $\widetilde{d\tau}(p, \tilde{x}) = (p, \tau(\tilde{x}))$, and, if we extend it to the associated vector bundles, we have that a section $s' \in \Gamma((\pi^{*}P_{\Pin^{\pm}}X) \times_{\rho} \mathbb{C}^{2^{n}})$ is the pull-back of a section on $X$ if and only if $\widetilde{d\tau}(s') = s'$.

Viceversa, let us start from $\tilde{X}$. We fix a pin$^{\pm}$ structure $\tilde{\xi}$ such that $\tilde{\xi} \simeq \tau^{*}\tilde{\xi}$ with $\widetilde{d\tau}^{2} = 1$. Then there are two natural vector space isomorphisms:
\begin{itemize}
	\item $\widetilde{d\tau}$-invariant sections of the associated bundle correspond to sections of the pin$^{\pm}$ structure $P_{\Pin^{\pm}}\tilde{X} \,/\, \widetilde{d\tau}$ on $X$;
	\item $(\widetilde{d\tau} \circ \gamma)$-invariant sections of the associated bundle correspond to sections of the pin$^{\pm}$ structure $P_{\Pin^{\pm}}\tilde{X} \,/\, (\widetilde{d\tau} \circ \gamma)$ on $X$.
\end{itemize}
In particular, $s = \widetilde{d\tau} \circ \eta_{\tau}(s)$ means that $s_{x} = \widetilde{d\tau}(s_{\tau(x)})$, while $s = \widetilde{d\tau} \circ \gamma \circ \eta_{\tau}(s)$ means that $s_{x} = -\widetilde{d\tau}(s_{\tau(x)})$, since the action of $\gamma$ corresponds to the multiplication by $-1 \in \Pin^{\pm}(n)$. We remark that if we want to describe invariance of pinors under general isometries, we must take into account the sign ambiguity discussed in the first section. In the present case, since we distinguish $\widetilde{d\tau}$ and $\widetilde{d\tau} \circ \gamma$ on the basis of the associated quotient on $X$, we fix this ambiguity. In this way we compensate the lack of injectivity of the map $\Phi$ between pin$^{\pm}$-structures on $X$ and $\tau$-invariant pin$^{\pm}$-structures on $\tilde{X}$, restoring the injectivity on pinors as sections of the associated vector bundles.

\subsection{Spinors and orientation-reversing isometries}\label{SpinorsOrientRev}

We now consider spinors instead of pinors on $\tilde{X}$. We have to correctly define the notion of $\tau$-invariance considering that $\tau$ reverses the orientation: in the definition represented by diagram \eqref{EquivalencePhiStar}, if $\varphi$ does not preserve the orientation, the differential $d\varphi$ does not have the same $\SO$-bundle for domain and codomain.\footnote{The two bundles could be isomorphic but not canonically, so we cannot think of $d\varphi$ as an automorphism anyway.} Thus in this case it does not make sense to speak about invariant spin structures. Fixing an orientation $u$ and a spin structure $\xi_{u}$ relative to $u$, we have seen that the vector bundles of spinors $q: S \rightarrow M$ and $q': S'\rightarrow M$ corresponding to $\xi_{u}$ and $\varphi^{*}\xi_{u}$ satisfy $S'_{x} = S_{\varphi^{-1}(x)}$. If we split the bundles into chiral sub-bundles $S = S^{+} \oplus S^{-}$ and $S' = {S'}^{+} \oplus {S'}^{-}$, we have that for $\varphi$ orientation-reversing ${S'}^{+}_{x} = S^{-}_{\varphi^{-1}(x)}$ and viceversa, i.e.\ the chiralities are reversed.\footnote{The reason for this is that the chirality element of the Clifford algebra of $TM$, which is the product of elements of an oriented orthonormal basis, becomes pointwise its own opposite if we change orientation (it is enough to multiply by $-1$ one of the vectors). So for $\xi_{u}$ and $\varphi^{*}\xi_{u}$ the chirality elements are opposite.} Thus, in order to define invariance, we must consider the case of different spin structures for the two chiralities, i.e.\ we must deal with ordered couples of spin structures.
\begin{Def} The \emph{bundle of spinors} associated to an ordered couple of spin structures $(\xi_{u}, \xi_{u'}')$ is the vector bundle $S_{\xi_{u},\xi_{u'}'} := S^{+}_{\xi_{u}} \oplus S^{-}_{\xi_{u'}'}$, where $S^{+}_{\xi_{u}}$ is the bundle of positive-chiral spinors with structure $\xi_{u}$ and $S^{-}_{\xi_{u'}'}$ is the bundle of negative-chiral spinors with structure $\xi_{u'}'$.
\end{Def}
\begin{Def} An ordered couple of spin structures $(\xi_{u}, \xi_{u'}')$ is \emph{oriented} if $u = u'$, i.e.\ if both spin structures lift the bundle of frames relative to the same orientation.
\end{Def}
We can now consider the pull-back of couples of spin structures.
\begin{Def} For $\varphi: M \rightarrow M$ an isometry and $(\xi_{u}, \xi_{u'}')$ an ordered couple of spin structures, we define:
\begin{itemize}
	\item for $\varphi$ orientation-preserving, $\varphi^{*}(\xi_{u}, \xi_{u'}') := (\varphi^{*}\xi_{u}, \varphi^{*}\xi_{u'}')$;
	\item for $\varphi$ orientation-reversing, $\varphi^{*}(\xi_{u}, \xi_{u'}') := (\varphi^{*}\xi_{u'}', \varphi^{*}\xi_{u})$.
\end{itemize}
We say that $(\xi_{u}, \xi_{u'}')$ is \emph{invariant under $\varphi$} if $\varphi^{*}(\xi_{u}, \xi_{u'}') \simeq (\xi_{u}, \xi_{u'}')$, where $\simeq$ means that they are componentwise equivalent.
\end{Def}
For $\varphi$ orientation-preserving, if $\xi_{u}$ is invariant then the couple $(\xi_{u}, \xi_{u})$ is invariant. If $\varphi$ is orientation-reversing, when the two components are equal the couple is never invariant; in order for a couple to be invariant in the latter case it must satisfy $\xi_{u'}' \simeq \varphi^{*}\xi_{u}$ and $\xi_{u} \simeq \varphi^{*}\xi_{u'}'$. There is a canonical representative $(\xi_{u}, \varphi^{*}\xi_{u})$, satisfying $(\varphi^{*})^{2}\xi_{u} \simeq \xi_{u}$, and diagram \eqref{EquivalencePhiStar} becomes:
	\[\xymatrix{
	P_{\Spin}M \ar[d]_{\xi_{u}} \ar@{.>}[r]^{\id, \gamma} & P_{\Spin}M \ar[d]^{\varphi^{*}\xi_{u}} \ar[dl]_{\varphi^{*}\xi_{u}} \\
	P_{\SO_{u}}M \ar[d]_{p} \ar[r]^{d\varphi} \ar[r]^{d\varphi} & P_{\SO_{u'}}M \ar[d]^{p'} \\
	M \ar[r]^{\varphi} & M
}\]
where $\gamma$ is the sheet exchange. However, we can canonically choose the identity as equivalence. The ambiguity is left in the choice of $\widetilde{d(\varphi^{2})}$ for the equivalence $(\varphi^{*})^{2}\xi_{u} \simeq \xi_{u}$. However, when $\varphi$ is an (orientation-reversing) involution, every couple of the form $(\xi_{u}, \varphi^{*}\xi_{u})$ is $\varphi$-invariant, and no ambiguity is left in the choice of equivalences.

\paragraph{}We now consider the invariance of spinors for $\varphi$-invariant couples of spin structures. In the orientation-preserving case, since the two elements of a couple are independent, the same considerations of the first paragraph about pinors apply separately for both elements of the couple. For the orientation-reversing case, we have the same ambiguity for couples $(\xi_{u}, \xi_{u'}')$ with $\xi_{u'}' \simeq \varphi^{*}\xi_{u}$ and $\xi_{u} \simeq \varphi^{*}\xi_{u'}'$. If, as discussed before, $\varphi$ is an involution and we choose the canonical representative $(\xi_{u}, \varphi^{*}\xi_{u})$, we do not have any ambiguity left, and we have a good notion of $\varphi$-invariant spinor. We call the involution $\tau$ instead of $\varphi$. In this case, the invariance acts as follows: let us consider a section $s \in \Gamma(S_{\xi_{u}, \tau^{*}\xi_{u}})$, i.e.\ a section $(s^{+}, s^{-}) \in \Gamma(S^{+}_{\xi_{u}} \oplus S^{-}_{\tau^{*}\xi_{u}})$. Then:
	\[s^{+}_{x} \in (S^{+}_{\xi_{u}})_{x} \qquad s^{-}_{x} \in (S^{-}_{\tau^{*}\xi_{u}})_{x} = (S^{+}_{\xi_{u}})_{\tau(x)}
\]
and we have the identity $(S^{+}_{\xi_{u}})_{x} \longrightarrow (S^{+}_{\xi_{u}})_{x} = (S^{-}_{\tau^{*}\xi_{u}})_{\tau(x)}$ inducing on sections the map $\eta_{\tau}: \Gamma(S^{+}_{\xi_{u}}) \rightarrow \Gamma(S^{-}_{\tau^{*}\xi_{u}})$. Thus we can ask $s^{-} = \eta_{\tau}(s^{+})$, i.e.\ $s^{-}_{x} = \eta_{\tau}(s^{+})_{x} = s^{+}_{\tau(x)}$. An invariant couple is thus of the form $(s^{+}, \eta_{\tau}(s^{+}))$, which corresponds to $(s^{+}_{x}, s^{+}_{\tau(x)})$.\footnote{We remark that given a generic couple $(s^{+}, s^{-})$ we can write it as $(\textstyle\frac{1}{2}(s^{+} + \eta(s^{-})) + \frac{1}{2}(s^{+} - \eta(s^{-})), \; \frac{1}{2}(s^{-} + \eta(s^{+})) + \frac{1}{2}(s^{-} - \eta(s^{+})))$ so that we can project it to an invariant couple by considering $(\textstyle\frac{1}{2}(s^{+} + \eta(s^{-})), \; \frac{1}{2}(s^{-} + \eta(s^{+})))$. This is coherent with closed unoriented superstring theory, where the quantum states surviving the projection are of the form $(\psi_{n} + \tilde{\psi}_{n})\vert 0 \rangle$.}

\paragraph{}We have thus seen that for any spin structure $\xi_{u}$ on $\tilde{X}$ the couple $(\xi_{u}, \tau^{*}\xi_{u})$ is $\tau$-invariant. Similarly, for spinors it is enough to choose a section $s$ realtive to $\xi_{u}$ to build in a unique way an invariant couple $(s^{+}_{x}, s^{+}_{\tau(x)})$. Thus, this definition of invariance for couples of spin structures does not entail any relationship with pin$^{\pm}$ structures on $X$.

\subsection{Spinors on the double cover and pinors}

In order to relate spinors $\tilde{X}$ and pinors on $X$ we need a condition more on the $\tau$-invariant couple $(\xi_{u}, \tau^{*}\xi_{u})$. Let us suppose that $\xi$ is a $\tau$-invariant pin$^{\pm}$ structure: then, if we restrict it to the spin structures $\xi_{u}$ and $\xi_{u'}$ corresponding to the two orientations $u$ and $u'$, we get isomorphisms $\xi_{u} \simeq \tau^{*}\xi_{u'}$ and $\xi_{u'} \simeq \tau^{*}\xi_{u}$ because the diagram:
	\[\xymatrix{
	P_{\Pin^{\pm}}\tilde{X} \ar[r]^{\widetilde{d\tau}} \ar[d]^{\xi} & P_{\Pin^{\pm}}\tilde{X} \ar[d]^{\xi} \ar[dl]_{\tau^{*}\xi} \\
	P_{O}\tilde{X} \ar[r]^{d\tau} & P_{O}\tilde{X}
}\]
restricts to:
	\[\xymatrix{
	P_{\Spin_{u}}\tilde{X} \ar[r]^{\widetilde{d\tau}} \ar[d]^{\xi_{u}} & P_{\Spin_{u'}}\tilde{X} \ar[d]^{\xi_{u'}} \ar[dl]_{\tau^{*}\xi_{u'}} & & P_{\Spin_{u'}}\tilde{X} \ar[r]^{\widetilde{d\tau}} \ar[d]^{\xi_{u'}} & P_{\Spin_{u}}\tilde{X} \ar[d]^{\xi_{u}} \ar[dl]_{\tau^{*}\xi_{u}} \\
	P_{SO_{u}}\tilde{X} \ar[r]^{d\tau} & P_{SO_{u'}}\tilde{X} & & P_{SO_{u'}}\tilde{X} \ar[r]^{d\tau} & P_{SO_{u}}\tilde{X}.
}\]
In particular, we get an isomorphism of couples $\widetilde{d\tau}: (\xi_{u}, \tau^{*}\xi_{u}) \overset{\simeq}\longrightarrow (\tau^{*}\xi_{u'}, \xi_{u'})$. We can now formulate the additional condition on a $\tau$-invariant couple of spin structures $(\xi_{u}, \tau^{*}\xi_{u})$, as defined in the previous subsection. We extend $\xi_{u}$ to a pin$^{\pm}$ structure $\xi$ via $P_{\Pin^{\pm}}\tilde{X} = P_{\Spin}\tilde{X} \times \Pin^{\pm}(n) \,/ \sim_{\Spin}$, where $(p, t) \sim_{\Spin} (ps, s^{-1}t)$ for $s \in \Spin(n)$. Then, reducing $\xi$ with respect to the other orientation we get $\xi_{u'}$, and we require that there is an isomorphism of couples:
\begin{equation}\label{DTauTilde}
	\widetilde{d\tau}: (\xi_{u}, \tau^{*}\xi_{u}) \overset{\simeq}\longrightarrow (\tau^{*}\xi_{u'}, \xi_{u'})
\end{equation}
satisfying $\widetilde{d\tau}^{2} = 1$. It easily follows that $\widetilde{d\tau}$ gives an equivalence of pin$^{\pm}$ structures between $\xi$ and $\tau^{*}\xi$, so that $\xi$ is the pull-back of a pin$^{\pm}$ structure on $X$.\footnote{In other words, there are two ways to relate spin structures relative to the two different orientations. The first one is the map $\xi_{u} \rightarrow \xi_{u'}$ obtained via the corresponding pin$^{\pm}$ structure as we have just explained, the second one is via $\tau^{*}$. We require that, for a fixed $\xi_{u}$, these two maps give the same value up to equivalence, and we also require that the equivalence squares to $1$.}

\paragraph{Remark:} This requirement depends upon whether we pass through pin$^{+}$ or pin$^{-}$ extensions. One may wonder whether the two ways lead to the same result. Actually it is true that, starting from $\xi_{u}$, we obtain the same structure $\xi_{u'}$ in both cases (we obtain the same result we would get reversing the orientation as described in \cite{KT}, without referring to pinors), but, as we will explicitely see for surfaces, the fact that $\widetilde{d\tau}^{2} = 1$ depends on the kind of pin structure we consider.

\paragraph{}We now analyze the behaviour of spinors as sections of the associated bundle. For pinors, fixing a pin$^{\pm}$ structure $\xi$ such that $\xi \simeq \tau^{*}\xi$ with $\widetilde{d\tau}^{2} = 1$, we can consider the couple of sections $(s,s)$ relative to the couple of pin$^{\pm}$ structures $(\xi, \tau^{*}\xi \simeq \xi)$. The section $s$ is supposed to satisfy $s = \widetilde{d\tau} \circ \eta_{\tau}(s)$, where the sign of $\widetilde{d\tau}$ is fixed by requiring that $P_{\Pin^{\pm},\xi}\tilde{X} \,/\, \widetilde{d\tau}$ is isomorphic to the pin$^{\pm}$ structure on $X$ we started from. Then we restrict to the \emph{oriented} couple of spin structures $(\xi_{u}, \tau^{*}\xi_{u'})$, so that the bundle of spinors is split in chiralities. Then we restrict $s$ to $S^{+}_{\xi_{u}}$ and $S^{-}_{\tau^{*}\xi_{u'}}$ obtaining $(s^{+}, s^{-})$. It follows that $s^{-} = \widetilde{d\tau} \circ \eta_{\tau}(s^{+})$, i.e.\footnote{We remark that $\eta_{\tau}$ reverses the chiralities with respect to $\xi$ and $\tau^{*}\xi$, but here $s^{+}$ is positive-chiral with respect to $\xi$, so $\eta_{\tau}(s^{+}_{x}) = s^{+}_{\tau(x)}$, as it is obvious from the fact that $\eta_{\tau}$ is the identity on the total space.} $s^{-}_{x} = \widetilde{d\tau}(s^{+}_{\tau(x)})$, so that we have a natural bijection between invariant pinors and couple of invariant spinors. If we consider $\widetilde{d\tau} \circ \gamma$, we get $s^{-} = \widetilde{d\tau} \circ \gamma \circ \eta_{\tau}(s^{+}) = -\widetilde{d\tau} \circ \eta_{\tau}(s^{+})$. Hence, as $\widetilde{d\tau}$-invariant couples are of the form $(s^{+}, \widetilde{d\tau} \circ \eta_{\tau}(s^{+}))$, similarly $(\widetilde{d\tau} \circ \gamma)$-invariant couples are of the form $(s^{+}, -\widetilde{d\tau} \circ \eta_{\tau}(s^{+}))$.\footnote{In particular, for a given couple $(s^{+}, t^{-})$, we have the two projectors $(\frac{1}{2}(s^{+} + \widetilde{d\tau} \circ \eta_{\tau}(t^{-}), \frac{1}{2}(t^{-} + \widetilde{d\tau} \circ \eta_{\tau}(s^{+})))$ and $(\frac{1}{2}(s^{+} - \widetilde{d\tau} \circ \eta_{\tau}(t^{-}), \frac{1}{2}(t^{-} - \widetilde{d\tau} \circ \eta_{\tau}(s^{+})))$.} Had we chosen the other orientation $u'$, we would have considered the couple $(\tau^{*}\xi_{u}, \xi_{u'})$, but since the sections of $\xi$ and $\tau^{*}\xi$ are the same as a subset of the total space, we would have obtained the same result. This is a consequence of the fact that for the couples $(\xi_{u}, \tau^{*}\xi_{u})$ and $(\xi_{u'}, \tau^{*}\xi_{u'})$ we have a canonical notion of $\tau$-invariant spinor.

\paragraph{}Summarizing:
\begin{enumerate}
	\item for a spin structure $\xi_{u}$ on $\tilde{X}$, the two couples $(\xi_{u}, \tau^{*}\xi_{u})$ and $(\xi_{u'}, \tau^{*}\xi_{u'})$ are alway $\tau$-invariant and there is a good notion of invariant spinor;
	\item to recover a relation with pinors on $\tilde{X}$, we require that there is an equivalence of couples $(\xi_{u}, \tau^{*}\xi_{u}) \simeq (\tau^{*}\xi_{u'}, \xi_{u'})$ via $\widetilde{d\tau}$;
	\item we fix an orientation $u$ of $\tilde{X}$ and consider the oriented couple $(\xi_{u}, \tau^{*}\xi_{u'})$, whose components are the first members of the two couples in point 2; it follows that the two components are equivalent via $\widetilde{d\tau}$ and we have a good notion of $\widetilde{d\tau}$-invariant couple of sections. These are the spinors we consider. Had we fixed the other orientation $u'$, we should consider the ordered couple $(\xi_{u'}, \tau^{*}\xi_{u})$, but thanks to point 1 we have a canonical bijection between the sections of this couple and the ones of the previous, thus the choice of the orientation is immaterial.
\end{enumerate}
Thus, as we have discussed in the introduction, the natural notion of $\tau$-invariance for couples of spin structures on $\tilde{X}$ does not entail any relation with the pin$^{\pm}$-structures on $X$. In order to implement such a relation we need a condition more, which contains anyway a reference to pin$^{\pm}$ structures on $\tilde{X}$, in particular it preserves the difference between pin$^{+}$ and pin$^{-}$ structures.

\section{Surfaces}

We show the explicit examples of pin structures on surfaces (cfr.\ \cite{DP1, DT}). All non-orientable surfaces can be obtained via connected sum of tori from the real projective plane or the Klein bottle. We use the following notations:
\begin{itemize}
	\item $\Sigma_{g}$ is the connected sum of $g$ tori;
	\item $N_{g,1}$ is the conntected sum of $\Sigma_{g}$ and the real projective plane $\mathbb{RP}^{2}$;
	\item $N_{g,2}$ is the conntected sum of $\Sigma_{g}$ and the Klein bottle $K^{2}$.
\end{itemize}

\subsection{One cross-cap}

We first consider the case of surfaces with one cross-cap, starting from the real projective plane $\mathbb{RP}^{2}$. Its orientable double covering is the 2-sphere $S^{2}$ with involution $\tau$ given by the reflection with respect to the center. In this case we have the following situation:
	\[H_{1}(S^{2}, \mathbb{Z}) = 0 \qquad H_{1}(\mathbb{RP}^{2}, \mathbb{Z}) = \mathbb{Z}_{2}
\]
so that the projection $\pi: S^{2} \rightarrow \mathbb{RP}^{2}$ induces a trivial immersion in homology, which is not surjective since there is a cycle in $\mathbb{RP}^{2}$ which does not lift to a cycle in $S^{2}$. Passing to $\mathbb{Z}_{2}$ coefficients we thus get:
	\[\begin{array}{lll}
	H_{1}(S^{2}, \mathbb{Z}_{2}) = 0 & & H_{1}(\mathbb{RP}^{2}, \mathbb{Z}_{2}) = \mathbb{Z}_{2}\\
	H^{1}(S^{2}, \mathbb{Z}_{2}) = 0 & & H^{1}(\mathbb{RP}^{2}, \mathbb{Z}_{2}) = \mathbb{Z}_{2}	
\end{array}\]
so that the induced maps are the zero maps:
	\[\pi_{*}: 0 \longrightarrow \mathbb{Z}_{2} \qquad \pi^{*}: \mathbb{Z}_{2} \longrightarrow 0
\]
and, in particular, the only non-trivial cohomology class, which corresponds to the element of $\Hom(H_{1}(X, \mathbb{Z}), \mathbb{Z}_{2})$ assigning $1$ to the class that does not lift, lies in the kernel of $\pi^{*}$.

All other surfaces $N_{g,1}$ with one number of cross-cap can be obtained adding $g$ tori to $\mathbb{RP}^{2}$ via connected sum. The double of $N_{g,1}$ is $\Sigma_{2g}$, namely the connected sum of $2g$-tori. In this case the situation is the following:
	\[H_{1}(\Sigma_{2g}, \mathbb{Z}) = \mathbb{Z}^{\oplus 4g} \qquad H_{1}(N_{g,1}, \mathbb{Z}) = \mathbb{Z}^{\oplus 2g} \oplus \mathbb{Z}_{2} \; .
\]
If we fix a canonical basis $\{a_{1}, b_{1}, \ldots, a_{2g}, b_{2g}\}$ of $H_{1}(\Sigma_{2g}, \mathbb{Z})$, the involution $\tau$ acts in such a way that $\tau_{*}(a_{i}) = a_{i+g}$ and $\tau_{*}(b_{i}) = b_{i+g}$ for $i = 1, \ldots, g$. There is a trivial cycle $c$ of which $\tau$ exchanges antipodal points: half of it is the lift of a representative of the $\mathbb{Z}_{2}$-generator of $N_{g,1}$ via $\pi: \Sigma_{2g} \rightarrow N_{g,1}$. Passing to $\mathbb{Z}_{2}$-coefficients we have:
	\[H_{1}(\Sigma_{2g}, \mathbb{Z}) = \mathbb{Z}_{2}^{\oplus 4g} \qquad H_{1}(N_{g,1}, \mathbb{Z}) = \mathbb{Z}_{2}^{\oplus 2g+1} \; .
\]
The push-forward $\pi_{*}$ sends $a_{i}$ and $a_{i+g}$ to the same class, similarly for $b_{i}$ and $b_{i+g}$. Instead of considering the $\mathbb{Z}_{2}$-reduction of the canonical basis, we consider the basis $\{a_{1}, b_{1}, \ldots,$ $a_{g}, b_{g}, a_{g+1} + a_{1}, b_{g+1} + b_{1}, \ldots, a_{2g} + a_{g}, b_{2g} + b_{g}\}$. It is still a basis since it can be obtained from the canonical one via the invertible $(4g \times 4g)$-matrix:
	\[\begin{bmatrix} I_{2g} & 0_{2g} \\ I_{2g} & I_{2g}
\end{bmatrix} \; .
\]
In this way we split $\mathbb{Z}_{2}^{\oplus 4g} = \mathbb{Z}_{2}^{\oplus 2g} \oplus \Ker \, \pi_{*}$, so that $\pi_{*}$ is injective on the first summand. Moreover, its image has index $2$ in $H_{1}(N_{g,1}, \mathbb{Z})$, since the only generator not lying in the image is the $\mathbb{Z}_{2}$-reduction of the $2$-torsion integral one. Thus we have the following picture for homology:
	\[\xymatrix{
	\mathbb{Z}_{2}^{\oplus 2g} \oplus (\mathbb{Z}_{2}^{\oplus 2g} = \Ker\, \pi_{*}) \ar@<-10ex>[d]^{\pi_{*}}_{\simeq} \; \,\\
	\mathbb{Z}_{2}^{\oplus 2g} \oplus (\mathbb{Z}_{2} \simeq \Coker\, \pi_{*})\;
}\]
which becomes in cohomology:
	\[\xymatrix{
	\mathbb{Z}_{2}^{\oplus 2g} \oplus (\mathbb{Z}_{2}^{\oplus 2g} \simeq \Coker\, \pi^{*}) \; \,\\
	\mathbb{Z}_{2}^{\oplus 2g} \oplus (\mathbb{Z}_{2} = \Ker\, \pi^{*})  \; . \quad\;\; \ar@<10ex>[u]^{\pi^{*}}_{\simeq}
}\]

\subsection{Two cross-caps}

We now consider the case of surfaces with two cross-caps, starting from the Klein bottle $K^{2} = N_{0,2}$. Its orientable double cover is the torus $T^{2} = \Sigma_{1}$ with involution $\tau$ defined in the following way: if we represent the torus as $\mathbb{C} / (2\pi\mathbb{Z} + 2\pi i\mathbb{Z})$, then we define
	\[\tau(z) = \overline{z} + \pi
\]
or equivalently in real coordinates $\tau(x, y) = (x + \pi, -y)$. We represent the torus as the square $[0, 2\pi] \times [0, 2\pi]$ with $(0, y) \sim (2\pi, y)$ and $(x, 0) \sim (x, 2\pi)$, and the Klein bottle as $[0, 2\pi] \times [0, 2\pi]$ with $(0, y) \sim (2\pi, y)$ and $(x, 0) \sim (2\pi - x, 2\pi)$. We call $a$ the loop $[0, 2\pi] \times \{0\}$ and $b$ the loop $\{0\} \times [0, 2\pi]$. We have that:
\begin{equation}\label{Pi1TK}
\begin{split}
	&\pi_{1}(T^{2}) = \langle \tilde{a}, \tilde{b} \,\vert\, \tilde{a}\tilde{b}\tilde{a}^{-1}\tilde{b}^{-1} = 1 \rangle\\
	&\pi_{1}(K^{2}) = \langle a, b \,\vert\, abab^{-1} = 1 \rangle
\end{split}
\end{equation}
The involution $\tau$ is the antipodal map of the $\tilde{a}$-generator, thus $(\tau_{*})_{\pi_{1}}[\tilde{a}] = [\tilde{a}]$, while it reflects the $\tilde{b}$-generator with respect to $y = \frac{1}{2}$ and apply the antipodal map, thus $(\tau_{*})_{\pi_{1}}[\tilde{b}] = [\tilde{b}]^{-1}$. The injective map induced by the projection is:
\begin{equation}\label{PiStarPi1}
\begin{split}
	(\pi_{*})_{\pi_{1}}: \; &\pi_{1}(T^{2}) \hookrightarrow \pi_{1}(K^{2})\\
	&(\pi_{*})_{\pi_{1}}(\tilde{a}) = b^{2}; \quad (\pi_{*})_{\pi_{1}}(\tilde{b}) = a \; .
\end{split}
\end{equation}
In homology, the abelianizations of \eqref{Pi1TK} are:
	\[H_{1}(T^{2}, \mathbb{Z}) = \mathbb{Z} \oplus \mathbb{Z} = \langle\!\langle \tilde{a}, \tilde{b} \rangle\!\rangle \qquad H_{1}(K^{2}, \mathbb{Z}) = \mathbb{Z} \oplus \mathbb{Z}_{2} = \langle\!\langle b, a \,\vert\, a^{2} = 1 \rangle\!\rangle
\]
(where $\langle\!\langle \,\cdot\, \rangle\!\rangle$ denotes the \emph{abelian} group with specified generators and relations) thus the map \eqref{PiStarPi1} becomes:
\begin{equation}\label{PiStar}
\begin{split}
	\pi_{*}: \; &\mathbb{Z} \oplus \mathbb{Z} \longrightarrow \mathbb{Z} \oplus \mathbb{Z}_{2}\\
	&\pi_{*}(1,0) = (2,0); \quad \pi_{*}(0,1) = (0,1) \; .
\end{split}
\end{equation}
Contrary to \eqref{PiStarPi1}, the map \eqref{PiStar} is not injetive any more, but its image has still index $2$ in the codomain, even if in this case the generator in $H_{1}(K^{2}, \mathbb{Z})$ not lifting to $H_{1}(T^{2}, \mathbb{Z})$ is the non-torsion one (i.e.\ the lifting of its representatives are now half of a non-trivial cycle of the covering, while for one-cross cap they were half of a trivial cycle). Passing to $\mathbb{Z}_{2}$ coefficients we get:
	\[\begin{array}{lll}
	H_{1}(T^{2}, \mathbb{Z}_{2}) = \mathbb{Z}_{2} \oplus \mathbb{Z}_{2} & & H_{1}(K^{2}, \mathbb{Z}_{2}) = \mathbb{Z}_{2} \oplus \mathbb{Z}_{2}\\
	H^{1}(T^{2}, \mathbb{Z}_{2}) = \mathbb{Z}_{2} \oplus \mathbb{Z}_{2} & & H^{1}(K^{2}, \mathbb{Z}_{2}) = \mathbb{Z}_{2}	\oplus \mathbb{Z}_{2}
\end{array}\]
so that the induced map in homology is:
	\[\begin{split}
	\pi_{*}:\;& H_{1}(T^{2}, \mathbb{Z}_{2}) \longrightarrow H_{1}(K^{2}, \mathbb{Z}_{2})\\
	&\pi_{*}(1,0) = 0 \qquad \pi_{*}(0,1) = (0,1)
\end{split}\]
For cohomology, we identify $(1,0) \in H^{1}(K^{2}, \mathbb{Z}_{2})$ with the functional $\varphi: H_{1}(K^{2}, \mathbb{Z}_{2}) \rightarrow \mathbb{Z}_{2}$ such that $\varphi(1,0) = 1$ and $\varphi(0,1) = 0$, and similarly for $(0,1) \in H^{1}(K^{2}, \mathbb{Z}_{2})$ with the functional $\psi$. Then $\pi^{*}(\varphi) = \varphi \circ \pi_{*}$ so that $\pi^{*}(\varphi)(1,0) = 0$ and $\pi^{*}(\varphi)(0,1) = 0$, while $\pi^{*}(\psi)(1,0) = 0$ and $\pi^{*}(\psi)(0,1) = 1$. Hence:
	\[\begin{split}
	\pi^{*}:\;& H^{1}(K^{2}, \mathbb{Z}_{2}) \longrightarrow H^{1}(T^{2}, \mathbb{Z}_{2})\\
	&\pi^{*}(1,0) = 0 \qquad \pi^{*}(0,1) = (0,1) \; .
\end{split}\]
and, in particular, the cohomology class which corresponds to the element of $\Hom(H_{1}(X, \mathbb{Z}), \mathbb{Z}_{2})$ assigning $1$ to the class that does not lift, lies in the kernel of $\pi^{*}$.

All other surfaces $N_{g,2}$ with two cross-caps can be obtained adding $g$ tori to $K^{2}$ via connected sum. The double of $N_{g,2}$ is $\Sigma_{2g+1}$, namely the connected sum of $2g$-tori. In this case the situation is the following:
	\[H_{1}(\Sigma_{2g+1}, \mathbb{Z}) = \mathbb{Z}^{\oplus 4g} \oplus \mathbb{Z} \oplus \mathbb{Z} \qquad H_{1}(N_{g,2}, \mathbb{Z}) = \mathbb{Z}^{\oplus 2g} \oplus \mathbb{Z} \oplus \mathbb{Z}_{2} \; .
\]
If we fix a canonical basis $\{a_{1}, b_{1}, \ldots, a_{2g+1}, b_{2g+1}\}$ of $H_{1}(\Sigma_{2g+1}, \mathbb{Z})$, the involution $\tau$ acts in such a way that:
\begin{itemize}
	\item $\tau_{*}(a_{i}) = a_{i+g}$ and $\tau_{*}(b_{i}) = b_{i+g}$ for $i = 1, \ldots, g$;
	\item $\tau_{*}(a_{2g+1}) = a_{2g+1}$ and $\tau_{*}(b_{2g+1}) = -b_{2g+1}$
\end{itemize}
and $\tau$ acts on two representatives $\tilde{a}$ of $a_{2g+1}$ and $\tilde{b}$ of $b_{2g+1}$ as in the case of the Klein bottle. Thus, half of $\tilde{a}$ is the lift of a representative of the last $\mathbb{Z}$-generator of $H^{1}(T^{2}, \mathbb{Z})$ via $\pi: \Sigma_{2g} \rightarrow N_{g,1}$, while $\tilde{b}$ is the lift of the $\mathbb{Z}_{2}$-generator. Passing to $\mathbb{Z}_{2}$-coefficients we have:
	\[H_{1}(\Sigma_{2g+1}, \mathbb{Z}_{2}) = \mathbb{Z}_{2}^{\oplus 4g} \oplus \mathbb{Z}_{2} \oplus \mathbb{Z}_{2} \qquad H_{1}(N_{g,2}, \mathbb{Z}_{2}) = \mathbb{Z}_{2}^{\oplus 2g} \oplus \mathbb{Z}_{2} \oplus \mathbb{Z}_{2} \; .
\]
The push-forward $\pi_{*}$ sends $a_{i}$ and $a_{i+g}$ to the same class, similarly for $b_{i}$ and $b_{i+g}$. As before, instead of considering the $\mathbb{Z}_{2}$-reduction of the canonical basis, we consider the basis $\{a_{1}, b_{1}, \ldots, a_{g}, b_{g},$ $a_{g+1} + a_{1}, b_{g+1} + b_{1}, \ldots, a_{2g} + a_{g}, b_{2g} + b_{g}, a_{2g+1}, b_{2g+1}\}$. In this way we split $\mathbb{Z}_{2}^{\oplus 4g}  \oplus \mathbb{Z}_{2} \oplus \mathbb{Z}_{2} = \mathbb{Z}_{2}^{\oplus 2g}  \oplus \mathbb{Z}_{2} \oplus \Ker \, \pi_{*}$, so that $\pi_{*}$ is injective on the first summand. Moreover, its image has index $2$ in $H_{1}(N_{g,2}, \mathbb{Z})$, since the only generator not lying in the image is the $\mathbb{Z}_{2}$-reduction of the $\mathbb{Z}$-factor of $K^{2}$. Thus we have the following picture:
	\[\xymatrix{
	\mathbb{Z}_{2}^{\oplus 2g} \oplus \mathbb{Z}_{2} \oplus (\mathbb{Z}_{2}^{\oplus 2g} \oplus \mathbb{Z}_{2} = \Ker\, \pi_{*}) \ar@<-12.5ex>[d]^{\pi_{*}}_{\simeq} \; \,\\
	\mathbb{Z}_{2}^{\oplus 2g} \oplus \mathbb{Z}_{2} \oplus (\mathbb{Z}_{2} \simeq \Coker\, \pi^{*}) \qquad\;\;
}\]
which becomes in cohomology:
	\[\xymatrix{
	\mathbb{Z}_{2}^{\oplus 2g} \oplus \mathbb{Z}_{2} \oplus (\mathbb{Z}_{2}^{\oplus 2g} \oplus \mathbb{Z}_{2} \simeq \Coker\, \pi^{*}) \; \,\\
	\mathbb{Z}_{2}^{\oplus 2g} \oplus \mathbb{Z}_{2} \oplus (\mathbb{Z}_{2} = \Ker\, \pi^{*}) \ar@<13ex>[u]^{\pi^{*}}_{\simeq} \;.\qquad\qquad\;
}\]

\subsection{Invariant structures on the sphere}

We think of the sphere $S^{2}$ as the Riemann sphere $\mathbb{CP}^{1}$, with two charts $U_{0} = \mathbb{CP}^{1} \setminus \{N\}$ and $U_{1} = \mathbb{CP}^{1} \setminus \{S\}$ and transition function $g_{01}(z) = -\frac{1}{z}$. The antipodal involution $\tau$ is specified each of the two charts\footnote{Note that $\tau$ commutes with $g_{01}$, that's why the expression is the same in both charts.} by $\tau(z) = -\frac{1}{\overline{z}}$. We compute its Jacobian to find the action $d\tau$ on the tangent bundle. In real cohordinates:
	\[\tau(x,y) = \Bigl( \frac{-x}{x^{2} + y^{2}}, \frac{-y}{x^{2} + y^{2}} \Bigr)
\]
so that the Jacobian becomes:
	\[J\tau(x, y) = \frac{1}{x^{2} + y^{2}} \begin{bmatrix} x^{2} - y^{2} & 2xy \\ 2xy & y^{2} - x^{2}
\end{bmatrix} \]
which, on the equator $\abs{z} = 1$ becomes the orthogonal matrix:
\begin{equation}\label{JacTau}
	J\tau(\cos \theta, \sin \theta) = \begin{bmatrix} \cos 2\theta & \sin 2\theta \\ \sin 2\theta & -\cos 2\theta
\end{bmatrix} \; .
\end{equation}
We now consider the sphere as the union of the two halves glued on the equator, so that we restrict both the charts $U_{0}$ and $U_{1}$ to the disc $\abs{z} \leq 1$, and we glue them via $g_{01}$. Now we consider the trivial spin structure for each of the two discs and we glue via a lift of $dg_{01}$ on $\abs{z} = 1$. On the equator of both charts the transformation \eqref{JacTau} is a reflection with respect to the real line generated by $(\cos \theta, \sin \theta)$, i.e.\ by $(-\sin \theta, \cos \theta)^{\bot}$. Thus, if we consider the point $(\cos \theta, \sin \theta) \in \mathbb{C} \simeq U_{0}$, we get $\tau(\cos \theta, \sin \theta) = -(\cos \theta, \sin \theta)$ and $d\tau_{(\cos \theta, \sin \theta)}$ acts on the tangent bundle as a rotation of $\pi$ along the equator composed with a reflection of the orthogonal direction. Hence its possible lifts to a $\Pin^{\pm}$-principal bundle are:
	\[\widetilde{d\tau}(\theta, p) = (\pi + \theta, \pm (-\sin \theta e_{1} + \cos \theta e_{2}) \cdot p) \; .
\]
Then $\widetilde{d\tau}^{2}$ is given by $(-\sin (\theta + \pi) e_{1} + \cos (\theta + \pi) e_{2})(-\sin \theta e_{1} + \cos \theta e_{2}) = (\sin \theta e_{1} - \cos \theta e_{2})(-\sin \theta e_{1} + \cos \theta e_{2}) = -\sin^{2}\theta e_{1}^{2} - \cos^{2}\theta e_{2}^{2}$. Thus we see that $\widetilde{d\tau}^{2} = 1$ if and only if $e_{1}^{2} = e_{2}^{2} = -1$, namely if the structure is $\Pin^{-}$: this shows that $\mathbb{RP}^{2} \simeq S^{2} \,/\, \tau$ has two pin$^{-}$ structures, lifting to the one of the sphere, but no pin$^{+}$ structures (compare with \cite{KT}).

\subsection{Invariant structures on the torus}

The torus has trivial tangent bundle $T^{2} \times \mathbb{R}^{2} \simeq S^{1} \times S^{1} \times \mathbb{R}^{2}$. The four inequivalent $\Spin$ or pin$^{\pm}$ structures can be all obtained from the trivial principal bundle $S^{1} \times S^{1} \times \Spin(2)$ or $S^{1} \times S^{1} \times \Pin^{\pm}(2)$ in the following way:
	\[\xymatrix{
	(\theta, \varphi, p') \ar[d]^{\tilde{\xi}_{0}} & & (\theta, \varphi, p') \ar[d]^{\tilde{\xi}_{1}} & & (\theta, \varphi, p') \ar[d]^{\tilde{\xi}_{2}} & & (\theta, \varphi, p') \ar[d]^{\tilde{\xi}_{3}}\\
	(\theta, \varphi, p)                           & & (\theta, \varphi, R_{\theta} \cdot p)          & & (\theta, \varphi, R_{\varphi} \cdot p)         & & (\theta, \varphi, R_{\varphi}R_{\theta} \cdot p)
}\]
where $R_{x}$ is the rotation by the angle $x$. To see that, e.g., the first two are not equivalent, we notice that we would need a map:
	\[\xymatrix{
	(\theta, \varphi, p') \ar[rr]^{\rho} \ar[dr]_{\tilde{\xi}_{0}} & & (\theta, \varphi, \tilde{R}_{-\theta}p') \ar[dl]^{\tilde{\xi}_{1}} \\
	& (\theta, \varphi, p)
}\]
for $\tilde{R}_{-\theta}$ a lift of $R_{-\theta}$ to $\Spin$ or $\Pin^{\pm}$. But in this way $\rho$ is not well defined, since for $\theta$ and $\theta + 2\pi$ we get two lifts differing by $-1$.

We now see that all these pin$^{\pm}$ structures are $\tau$-invariant, where $\tau$ is the involution giving the Klein bottle, namely $\tau(\theta, \varphi) = (\theta + \pi, -\varphi)$. On the tangent frame bundle we have the action $d\tau(\theta, \varphi, p) = (\theta + \pi, -\varphi, j_{2}p)$ where $j_{2}$ is the reflection along $e_{2}^{\bot}$, i.e.\ $(x, y) \rightarrow (x, -y)$. The equivalence between $\tilde{\xi}_{0}$ and $\tau^{*}\tilde{\xi}_{0}$ is given by the following diagram:
	\[\xymatrix{
	(\theta, \varphi, p') \ar[r]^{\widetilde{d\tau} \qquad} \ar[d]_{\tilde{\xi}_{0}} & (\theta + \pi, -\varphi, e_{2} \cdot p') \ar[d]^{\tilde{\xi}_{0}} \ar[dl]_{\tau^{*}\tilde{\xi}_{0}} \\
	(\theta, \varphi, p) \ar[r]^{d\tau \quad\;\;} & (\theta + \pi, -\varphi, j_{2}p)
}\]
or equivalently by $\widetilde{d\tau} \circ \gamma$ which can be obtained by choosing $-e_{2}$. Here we see that for the $\Pin^{+}$-structure, since $e_{2}^{2} = 1$, we get $\widetilde{d\tau}^{2} = 1$, while for the $\Pin^{-}$-structure we get $\widetilde{d\tau}^{2} = -1$. Thus, only the $\Pin^{+}$-structure is the pull-back of a $\Pin^{+}$-structure of $K^{2}$. For $\tilde{\xi}_{1}$:
	\[\xymatrix{
	(\theta, \varphi, p') \ar[r]^{\widetilde{d\tau} \qquad\qquad} \ar[d]_{\tilde{\xi}_{1}} & (\theta + \pi, -\varphi, \tilde{R}_{-\theta-\pi} e_{2} \tilde{R}_{\theta} p') \ar[d]^{\tilde{\xi}_{1}} \ar[dl]_{\tau^{*}\tilde{\xi}_{1}} \\
	(\theta, \varphi, R_{\theta} p) \ar[r]^{d\tau \qquad} & (\theta + \pi, -\varphi, j_{2} R_{\theta} p)
}\]
and $\widetilde{d\tau}$ is well-defined since with the shift $\theta \rightarrow \theta + 2\pi$ we get a minus sign in both liftings of the rotations. Then $\widetilde{d\tau}^{2} = \tilde{R}_{-(\theta+\pi)-\pi} e_{2} \tilde{R}_{\theta+\pi} \tilde{R}_{-\theta-\pi} e_{2} \tilde{R}_{\theta} = \tilde{R}_{-2\pi}e_{2}^{2} = -e_{2}^{2}$, thus we get opposite results with respect to $\tilde{\xi}_{0}$. For $\tilde{\xi}_{2}$:
	\[\xymatrix{
	(\theta, \varphi, p') \ar[r]^{\widetilde{d\tau} \qquad\qquad\quad} \ar[d]_{\tilde{\xi}_{2}} & (\theta + \pi, -\varphi, \tilde{R}_{\varphi} e_{2} \tilde{R}_{\varphi} p') \ar[d]^{\tilde{\xi}_{2}} \ar[dl]_{\tau^{*}\tilde{\xi}_{2}} \\
	(\theta, \varphi, R_{\varphi} p) \ar[r]^{d\tau \qquad} & (\theta + \pi, -\varphi, j_{2} R_{\varphi} p)
}\]
and $\widetilde{d\tau}$ is well-defined since with the shift $\theta \rightarrow \theta + 2\pi$ we get a minus sign in both liftings of the rotations. Then $\widetilde{d\tau}^{2} = \tilde{R}_{-\varphi} e_{2} \tilde{R}_{-\varphi} \tilde{R}_{\varphi} e_{2} \tilde{R}_{\varphi} = e_{2}^{2}$, thus we get the same results of $\tilde{\xi}_{0}$. It is clear that $\tilde{\xi}_{3}$ behaves as $\tilde{\xi}_{1}$.

\paragraph{}If we consider pinors and spinors as sections of the associated vector bundles, for $\tilde{\xi}_{0}$ the condition on pinors is $s_{(\theta, \varphi)} = e_{2} \cdot s_{(\theta + \pi, -\varphi)}$, while for spinors we consider the couple $(s^{+}, s^{-})$ with structure $((\tilde{\xi}_{0})_{u}, \tau_{*}(\tilde{\xi}_{0})_{u'}))$ where $s^{+}$ is free and $s^{-}_{(\theta, \varphi)} = e_{2} \cdot s^{+}_{(\theta + \pi, -\varphi)}$. Similar conditions for the other spin structures.

\section{Manifolds with boundary}

We now want to give the analogous description in the case of unorientable manifolds \emph{with boundary}. We start with a brief recall of the well-known case of spinors on orientable manifolds with boundary, in order to extend it to pinors and discuss the non-orientable case.

\subsection{Orientable manifolds with boundary}

Let $X$ be an orientable manifold of dimension $2n$ with boundary $\partial X$, and let us consider its double $X^{d}$ obtained considering two disjoint copies of $X$ and identifying the corresponding boundary points. We mark one of the two copies considering an embedding $i: X \rightarrow X^{d}$. In this way, an orienation of $X^{d}$ induces an orientation of $X$ and the opposite one on the other copy. We have a natural orientation-reversing involution $\tau$ identifying corresponding points of the two copies, which is \emph{not} a double covering since the boundary points are fixed.

\paragraph{}\textbf{Remark:} We have a natural projection $\pi: X^{d} \longrightarrow X \simeq X^{d} \,/\, \tau$, but it is in general not smooth, since on a local curve orthogonal to a boundary point the behaviour of $\tau$ and $\pi$ is of the form $\tau(x) = -x$ and $\pi(x) = \abs{x}$. This is why in the open case it is more natural to deal with the immersion $i: X \rightarrow X^{d}$ which has no analogue in the closed non-orientable case.

\paragraph{}We consider on $X$ couples of spin structure $(\xi, \xi')$ with an isomorphism $\theta: \xi \vert_{\partial X} \rightarrow \xi' \vert_{\partial X}$, where the restriction is obtained in the following way: we consider the immersion $P_{O}(\partial X) \subset P_{O}X$ sending a basis $\{e_{1}, \ldots e_{2n-1}\}$ of $T_{x}(\partial X)$ to the basis $\{e_{1}, \ldots e_{2n-1}, u\}$ of $T_{x}X$ where $u$ is the outward orthogonal unit vector. We consider two triples $(\xi, \xi', \theta)$ and $(\eta, \eta', \varphi)$ equivalent if there exist equivalences $\rho_{1}: \xi \rightarrow \eta$ and $\rho_{2}: \xi' \rightarrow \eta'$ such that $\rho_{2}\vert_{\partial X}^{-1} \circ \varphi \circ \rho_{1}\vert_{\partial X} = \theta$. On each connected component $Y \subset \partial X$, there are two possibilities for $\theta\vert_{Y}$ linked by $\gamma$. An overall change from $\theta$ to $\theta \circ \gamma$ is irrelevant since $(\xi, \xi', \theta) \simeq (\xi, \xi', \theta \circ \gamma)$ via $\rho_{1} = \id$ and $\rho_{2} = \gamma$; instead, the separate restrictions determined by $\theta$ are meaningful, thus we must fix all of them except one. Let us show that equivalence classes of such triples $(\xi, \xi', \theta)$ correspond bijectively to equivalence classes of spin structures $\tilde{\xi}$ on $X^{d}$ associated only to positive chirality.\\
\textbf{From $X$ to $X^{d}$:} given $(\xi, \xi', \theta)$ we define $\tilde{\xi}\vert_{X} := \xi$ and $\tilde{\xi}\vert_{X^{d}\setminus \Int(X)} := \xi'$, and we glue them on $\partial X$ via the isomorphism $\theta$. We call such a spin structure $\xi \cup_{\theta} \xi'$. We can always restrict ourselves to the case $\theta = \id$ by considering $\bigl( (\xi \cup_{\theta} \xi')\vert_{X}, \tau^{*}((\xi \cup_{\theta} \xi')\vert_{X^{d} \setminus \Int(X)}), \id \bigr)$. \\
\textbf{From $X^{d}$ to $X$:} given $\tilde{\xi}$ we define $\xi := \tilde{\xi}\vert_{X}$ and $\xi' := \tau^{*}(\tilde{\xi}\vert_{X^{d}\setminus \Int(X)})$. The isomorphism $\theta$ is the identity on $\tilde{\xi}\vert_{\partial X}$.

\paragraph{}A few comments are in order. When we define $\xi' := \tau^{*}(\tilde{\xi}\vert_{X^{d}\setminus \Int(X)})$, the set $X^{d}\setminus \Int(X)$ is a copy of $X$ oriented in the opposite way, but $\tau$ recovers the original orientation, thus it reverses the chirality. In particular, considering spinors as sections of the associated vector bundles, we have a section $\tilde{s}^{+}$ on $X^{d}$ and a couple $(s^{+}, s^{-})$ on $X$, with the condition $\tilde{s}^{+}_{x} = s^{+}_{x}$ for $x \in X \setminus \partial X$ and $\tilde{s}^{+}_{x} = s^{-}_{\tau(x)}$ for $x \in X^{d} \setminus \Int(X)$, while on the boundary we consider $s^{+}_{x}$ on one copy, $s^{-}_{x}$ on the other and we glue them on $\partial X$ via $\theta: S^{+}_{x} \rightarrow S^{-}_{x}$ extended to the vector bundles. Of course we could also choose negative chirality on the double, by exchanging the roles of $\xi$ and $\xi'$.

When we double a manifold, we can create new cycles. Let us think of the case of a cylinder whose double is a torus. Two structures $\tilde{\xi}_{1}$ and $\tilde{\xi}_{2}$ on the double can differ by the holonomy of the spin connection along one of those cycles: the corresponding couples $(\xi_{1}, \xi_{1}', \theta_{1})$ and $(\xi_{2}, \xi_{2}', \theta_{2})$ will verify $\xi_{1} \simeq \xi_{2}$ and $\xi_{1}' \simeq \xi_{2}'$, but the difference can be read in $\theta_{1}$ and $\theta_{2}$: if we fix the same representative bundles for $(\xi_{1}, \xi_{2})$ and $(\xi_{1}', \xi_{2}')$, then $\theta_{1}$ and $\theta_{2}$ will differ by a $-1$ in one of the two boundary components intersecting the involved half-cycle (which of the two boundary components depends on the overall sign).

\paragraph{}In this section we have not referred so far to pinors, since an orientation was fixed both for $X$ and $X^{d}$ and there was no reason to relate it to the other orientation. For later use we need however to  consider a pin$^{\pm}$ structure on $X$, forgetting the orientation. In this case we do not consider a couple of pin$^{\pm}$ structures since we have no chirality, but we consider couples $(\xi, \theta)$ where $\theta: \xi\vert_{\partial X} \rightarrow \xi\vert_{\partial X}$ is an automorphism. Then we can glue two copies of $\xi$ on $X^{d}$ to $\xi \cup_{\theta} \xi$. For every connected component $Y \subset \partial X$, since $\theta\vert_{Y}$ lifts the identity of the tangent bundle of $Y$, it must be the identity or $\gamma$. Viceversa, if we have a pin$^{\pm}$ structure $\tilde{\xi}$ on $X^{d}$, then $\tilde{\xi}$ is equivalent to $\tilde{\xi}\vert_{X} \cup_{\id} \tau_{*}(\tilde{\xi}\vert_{X^{d} \setminus \Int(X)})$. If there exists an isomorphism $\widetilde{d\tau}: \tilde{\xi}\vert_{X} \overset{\simeq}\longrightarrow \tau_{*}(\tilde{\xi}\vert_{X^{d} \setminus \Int(X)})$, we restrict the latter to $\widetilde{d\tau}\vert_{\partial X}: \tilde{\xi}\vert_{\partial X} \overset{\simeq}\longrightarrow \tilde{\xi}\vert_{\partial X}$ and if we apply $\widetilde{d\tau}^{-1}$ to the second component, we obtain $\tilde{\xi}\vert_{X} \cup_{\widetilde{d\tau}\vert_{\partial X}^{-1}} \tilde{\xi}\vert_{X}$ (we can freely choose $\widetilde{d\tau}$ or $\widetilde{d\tau} \circ \gamma$ since they differ by an overall sign, thus we can suppose $\widetilde{d\tau}^{-1} = \widetilde{d\tau}$). On sections, we just ask $s_{x} = \widetilde{d\tau}(s_{\tau(x)})$. Thus we have an equivalence of categories:
	\[\left\{ \begin{array}{ll}
	(\xi, \theta) \,: & \xi \textnormal{ pin$^{\pm}$ structure on $X$}\\
	& \theta: \xi\vert_{\partial X} \overset{\simeq}\longrightarrow \xi\vert_{\partial X}
\end{array} \right\} \longleftrightarrow \left\{ \begin{array}{ll}
	\tilde{\xi} \,: & \tilde{\xi} \textnormal{ pin$^{\pm}$ structure on $X^{d}$ s.t.}\\
	& \exists \, \widetilde{d\tau}: \tilde{\xi}\vert_{X} \overset{\simeq}\longrightarrow \tau^{*}(\tilde{\xi}\vert_{X^{d} \setminus \Int(X)})
\end{array} \right\} \; .
\]

\paragraph{}Here are some remarks about this picture. The condition of the existence of $\widetilde{d\tau}: \tilde{\xi}\vert_{X} \simeq \tau^{*}\tilde{\xi}\vert_{X^{d} \setminus \Int(X)}$ has no analogue for spin structures since for that case we considered only the positive chirality on $X^{d}$. Should we consider also the negative chirality, the couple of spin structures should satisfy this condition. The same for sections. In particular, with spinors we have on $X$ the freedom of choosing positive and negative chiralities, with the condition that they must be isomorphic at the boundary. With pinors, which are extendable to the non-orientable case, there are no distinction between chiralites: this is the analogue of considering the same chirality for positive and negative spinors. In this case, for spin structures we should consider triples $(\xi, \xi, \theta)$, corresponding to the couples $(\xi, \theta)$ for pinors. If we start from the double, in the spin case we can freely choose $\tilde{\xi}$ for positive-chiral spinors, then on $X$ we have $\tilde{\xi}\vert_{X}$ for positive and $\tau_{*}(\tilde{\xi}\vert_{X^{d} \setminus \Int(X)})$ for negative ones: they are not in general isomorphic, but they coincide on the boundary since $\tau\vert_{\partial X}$ is the identity. For pinors (or spinors considering both chiralities) we recover the fact not to have only positive chirality by asking that there is an isomorphism $\widetilde{d\tau}$ between $\tau^{*}(\tilde{\xi}\vert_{X^{d} \setminus \Int(X)})$ and $\tilde{\xi}\vert_{X}$, but we do not ask that such isomorphism restricts to the identity on the boundary. Thus, if we glue the two pieces with the identity we have an isomorphism which is not in general a restriction of a global one as $\widetilde{d\tau}$: applying $\widetilde{d\tau}^{-1}$ to both members we obtain $\widetilde{d\tau}^{-1}$ itself as gluing isomorphism, but now both the members are equal and the identity between them still does not restrict to $\widetilde{d\tau}^{-1}$. Thus, we have a generic isomorphism at the boundary, not necessarily the restriction of a global one.

\subsection{Unorientable manifolds with boundary}

Let $X$ be an unorientable manifold with boundary. Then we can consider the diagram:
\begin{equation}\label{DiagramCoverings}
	\xymatrix{
	& (\tilde{X}, \tau_{1}) \ar@{->>}[dl]_{\pi_{1}} & \\
	X & & (\tilde{X}^{d}, \tau_{3}, \tau_{4}) \;. \ar@{->>}[ul]_{\pi_{3}} \ar@{->>}[dl]^{\pi_{4}} \\
	& (X^{d}, \tau_{2}) \ar@{->>}[ul]^{\pi_{2}} &
}
\end{equation}
We remark that we have immersions $X \subset X^{d}$ and $\tilde{X} \subset \tilde{X}^{d}$, while $\pi_{1}$ and $\pi_{4}$ are double coverings. In particular $\tau_{2}$ and $\tau_{3}$ have fixed points while $\tau_{1}$ and $\tau_{4}$ do not. By the definition of $\tilde{X}$ and $X^{d}$ with the relevant involutions we easily get the following properties:
\begin{itemize}
	\item $\pi_{1} \circ \pi_{3} = \pi_{2} \circ \pi_{4}$;
	\item $\pi_{4} \vert_{\tilde{X}} = \pi_{1}$ and $\tau_{4} \vert_{\tilde{X}} = \tau_{1}$;
	\item $\tau_{3} \circ \tau_{4} = \tau_{4} \circ \tau_{3}$.
\end{itemize}
As for the open oriented case, we must fix a couple $(\xi, \theta)$ with $\xi$ a pin$^{\pm}$ structure on $X$ and $\theta: \xi\vert_{\partial X} \rightarrow \xi\vert_{\partial X}$ an automorphism. To estabilish a correspondence with pin$^{\pm}$ structures on $\tilde{X}^{d}$, we can follow the upper or the lower paths of diagram \eqref{DiagramCoverings}. If we follow the lower path, we consider $\xi^{d(\theta)} := \xi \cup_{\theta} \xi$ on $X^{d}$, then we pull it back to $\pi_{4}^{*}(\xi^{d(\theta)})$. Otherwise, following the upper path, we first pull back $\xi$ to $\pi_{1}^{*}\xi$ as in the closed case, so that $\xi\vert_{\partial X}$ pulls-back to $(\pi_{1}^{*}\xi)\vert_{\partial \tilde{X}}$ and the morphism $\theta$ pulls back to a morphism $\pi_{1}^{*}\theta: (\pi_{1}^{*}\xi)\vert_{\partial \tilde{X}} \rightarrow (\pi_{1}^{*}\xi)\vert_{\partial \tilde{X}}$. Then we double $\pi_{1}^{*}\xi$ on $\tilde{X}^{d}$ putting it on both copies of $\tilde{X}$ and using $\pi_{1}^{*}\theta$ as the isomorphism on $\partial \tilde{X}$, i.e.\ we consider $\pi_{1}^{*}\xi \cup_{\pi_{1}^{*}\theta} \pi_{1}^{*}\xi$, which we call $(\pi_{1}^{*}\xi)^{d(\pi_{1}^{*}\theta)}$. The two results are the same, in fact $(\pi_{4}^{*}(\xi^{d(\theta)}))\vert_{\tilde{X}} = (\pi_{4}\vert_{\tilde{X}})^{*}(\xi^{d(\theta)}\vert_{X}) = \pi_{1}^{*}(\xi) = (\pi_{1}^{*}\xi)^{d(\pi_{1}^{*}\theta)}\vert_{\tilde{X}}$, and the same for the other half of $\tilde{X}^{d}$ and for the isomorphism $\theta$. Considering sections of the associated vector bundles of pinors, since under pull-back of pin$^{\pm}$ structures we pull-back also sections and under doubling we ask invariance of the sections, we obtain sections $s \in \Gamma(P_{\Pin^{\pm}, (\pi_{1}^{*}\xi)^{d(\pi_{1}^{*}\theta)}}(\tilde{X}^{d}) \times_{\rho} \mathbb{C}^{2^{n}})$, such that $s_{x} = s_{\tau_{3}(x)} = s_{\tau_{4}(x)} = s_{\tau_{3}\tau_{4}(x)}$. Here we do not have $\widetilde{d\tau_{3}}$ and $\widetilde{d\tau_{4}}$ since we are working with explicit pull-backs.

Viceversa, if we are given a pin$^{\pm}$ structure $\xi'$ on $\tilde{X}^{d}$, such that there exists $\widetilde{d\tau_{3}}: \xi'\vert_{\tilde{X}} \simeq (\tau_{3})_{*}(\xi'\vert_{\tilde{X}^{d} \setminus \Int(\tilde{X})})$ restricting to the boundary, and $\widetilde{d\tau_{4}}: \xi' \overset{\simeq}\longrightarrow (\tau_{4})_{*}\xi'$ with $\widetilde{d\tau_{4}}^{2} = 1$, then we can find a pin$^{\pm}$ structure on $X$ such that $\xi' \simeq (\pi_{1}^{*}\xi)^{d(\theta)}$. We can find it using the two paths of the diagram. If we follow the upper path, we consider the couple $(\xi'\vert_{\tilde{X}}, \id)$ where $\id: \xi'\vert_{\partial \tilde{X}} \rightarrow \xi'\vert_{\partial \tilde{X}}$ is the restriction of $\widetilde{d\tau_{3}}$. Then, since $\tau_{4} \vert_{\tilde{X}} = \tau_{1}$, if follows that $\xi'\vert_{\tilde{X}}$ is $\tau_{1}$-invariant with $\widetilde{d\tau_{1}}^{2} = 1$, thus we can consider $\xi = (\xi'\vert_{\tilde{X}}) \,/\, \widetilde{d\tau_{1}}$ as in the closed case. For sections on $\tilde{X}^{d}$, we must ask $s_{x} = \widetilde{d\tau_{4}}(s_{\tau_{4}(x)}) = \widetilde{d\tau_{3}}(s_{\tau_{3}(x)})$. If we follow the lower path of the diagram, we first quotient by $\widetilde{d\tau_{4}}$ and then we use the projection of $\widetilde{d\tau_{3}}$ to $X^{d}$ by $\pi_{4}$.

\paragraph{}Given the invariant structure $\xi'$, we can consider the couple of spin structures $(\xi'_{u}, (\tau_{4})_{*}\xi'_{u'})$ and the relative spinors $(s^{+}, s^{-})$: in this way the conditions become $s^{-}_{x} = \widetilde{d\tau_{4}}(s^{+}_{\tau_{4}(x)})$ and $s^{-}_{x} = \widetilde{d\tau_{3}}(s^{+}_{\tau_{3}(x)})$. Thus, $s^{-}$ is completely determined by $s^{+}$, and we have one condition $s^{+}_{x} = \widetilde{d\tau_{4}}\circ \widetilde{d\tau_{3}} (s^{+}_{\tau_{3}\tau_{4}(x)})$ (necessarily $\widetilde{d\tau_{3}}^{2} = 1$ and $\widetilde{d\tau_{3}} \circ \widetilde{d\tau_{4}} = \widetilde{d\tau_{4}} \circ \widetilde{d\tau_{3}}$).

If we follow the upper path starting from $\xi$ on $X$, instead of $\pi_{1}^{*}\xi$ we can directly consider a pin$^{\pm}$ structure $\xi$ with an isomorphism $\widetilde{d\tau_{1}}: \xi \rightarrow (\tau_{1})_{*}\xi$, such that $\widetilde{d\tau_{1}}^{2} = 1$, and the corresponding couple $(\xi_{u}, (\tau_{1})_{*}\xi_{u'})$. If we double such a couple to a spin structure on $\tilde{X}^{d}$ via $\theta = \widetilde{d\tau_{1}}\vert_{\partial X}$, we obtain exactly the structure of positive-chiral spinors obtained from $\xi'$ which is the double of $\xi$.

\paragraph{}We can also consider the orientation-\emph{preserving} involution $\tau_{34} = \tau_{3} \circ \tau_{4}$ on $\tilde{X}^{d}$. We now show that $X' = \tilde{X}^{d} \,/\, \tau_{34}$ is an oriented and closed manifold with an orientation-reversing involution $\tau'$ such that $X' \,/\, \tau' \simeq X$. In fact, $\tau_{34}$ has no fixed points: $\tau_{34}(x) = x$ is equivalent to $\tau_{3}(x) = \tau_{4}(x)$, but if $x \notin \partial \tilde{X} \subset \tilde{X}^{d}$, then $\tau_{3}$ maps it to a point of the other copy of $\tilde{X}$, while $\tau_{4}$ exchanges the sheets of the covering of the same copy of $X$; instead, if $x \in \partial \tilde{X} \subset \tilde{X}^{d}$, then $\tau_{3}(x) = x$ while $\tau_{4}$ has no fixed points. Therefore $\tau_{3}(x) = \tau_{4}(x)$ is impossible. Hence $X' = \tilde{X}^{d} \,/\, \tau_{34}$ is a smooth closed orientable manifolds double-covered by $\tilde{X}^{d}$. Then $\tau_{3}$ and $\tau_{4}$ projects at the quotient to the same involution $\tau'$. We can thus complete the diagram:
	\[\xymatrix{
	& (\tilde{X}, \tau_{1}) \ar@{->>}[dl]_{\pi_{1}} & \\
	X & (X', \tau') \ar@{->>}[l]_{\pi'} & (\tilde{X}^{d}, \tau_{3}, \tau_{4}) \;. \ar@{->>}[ul]_{\pi_{3}} \ar@{->>}[dl]^{\pi_{4}} \ar@{->>}[l]_{\pi_{34}} \\
	& (X^{d}, \tau_{2}). \ar@{->>}[ul]^{\pi_{2}} &
}\]
The previous picture is analogous to considering a pin$^{\pm}$ structure $\xi'$ on $X'$ which is $\tau'$-invariant with $\widetilde{d\tau'}^{2} = 1$. Via $\pi_{34}$ we pull-back it to a structure on $\tilde{X}^{d}$ satisfying the previous requirements.

\subsection{Moebius strip}

We now study as an example pinors on the Moebius strip. In this case diagram \eqref{DiagramCoverings} becomes (calling $\Cyl$ the finite cylinder or annulus and $M^{2}$ the Moebius strip):
	\[\xymatrix{
	& (\Cyl, \tau_{1}) \ar@{->>}[dl]_{\pi_{1}} & \\
	M^{2} & (T^{2}, \tau') \ar@{->>}[l]_{\pi'} & (T^{2}, \tau_{3}, \tau_{4}) \;. \ar@{->>}[ul]_{\pi_{3}} \ar@{->>}[dl]^{\pi_{4}} \ar@{->>}[l]_{\pi_{34}} \\
	& (K^{2}, \tau_{2}). \ar@{->>}[ul]^{\pi_{2}} &
}\]
with the involutions we now describe. We represent all the four surfaces involved as the square $[0, 2\pi] \times [0, 2\pi]$ with suitable identifications on the edges. In particular, for $M^{2}$ we identify $(0, y) \sim (2\pi, 2\pi - y)$, for $\Cyl$ $(0, y) \sim (2\pi, y)$, for $T^{2}$ $(0, y) \sim (2\pi, y)$ and $(x, 0) \sim (x, 2\pi)$, and for $K^{2}$ $(0, y) \sim (2\pi, y)$ and $(x, 0) \sim (2\pi - x, 2\pi)$. When two edges are identified with the same direction (\emph{and only in this case}), we think of the orthogonal coordinate as a $2\pi$-periodical coordinate $\mathbb{R} \,/\, 2\pi \mathbb{Z}$. With these conventions a possible choiche of involutions is:
	\[\begin{array}{lll}
	\tau_{1}(x,y) = (x + \pi, 2\pi - y) & & \tau_{2}(x,y) = (y - x, y) \\
	\tau_{3}(x,y) = (-x, y) & & \tau_{4}(x,y) = (\pi - x, y + \pi) \; .
	\end{array}
\]
We now analyze pin$^{\pm}$ structures on $T^{2}$. We can prove as before that they are all $\tau_{4}$-invariant. In the $(\theta, \varphi)$-coordinates $\tau_{4}$ becomes $\tau_{4}(\theta, \varphi) = (-\theta + \pi, \varphi + \pi)$. Then:
	\[\xymatrix{
	(\theta, \varphi, p') \ar[r]^{\widetilde{d\tau_{4}} \qquad\quad} \ar[d]_{\tilde{\xi}_{0}} & (-\theta + \pi, \varphi + \pi, e_{1} \cdot p') \ar[d]^{\tilde{\xi}_{0}} \ar[dl]_{(\tau_{4})^{*}\tilde{\xi}_{0}} \\
	(\theta, \varphi, p) \ar[r]^{d\tau_{4} \qquad\;} & (-\theta + \pi, \varphi + \pi, j_{1}p)
}\]
	\[\xymatrix{
	(\theta, \varphi, p') \ar[r]^{\widetilde{d\tau_{4}\;} \qquad\qquad\quad} \ar[d]_{\tilde{\xi}_{1}} & (-\theta + \pi, \varphi + \pi, \tilde{R}_{\theta-\pi} \cdot e_{1} \cdot \tilde{R}_{\theta} \cdot p') \ar[d]^{\tilde{\xi}_{1}} \ar[dl]_{(\tau_{4})^{*}\tilde{\xi}_{1}} \\
	(\theta, \varphi, R_{\theta} \cdot p) \ar[r]^{d\tau_{4} \qquad\quad} & (-\theta + \pi, \varphi + \pi, j_{1} \cdot R_{\theta} \cdot p)
}\]
	\[\xymatrix{
	(\theta, \varphi, p') \ar[r]^{\widetilde{d\tau_{4}} \qquad\qquad\qquad} \ar[d]_{\tilde{\xi}_{2}} & (-\theta + \pi, \varphi + \pi, \tilde{R}_{-\varphi-\pi} \cdot e_{1} \cdot \tilde{R}_{\varphi} \cdot p') \ar[d]^{\tilde{\xi}_{2}} \ar[dl]_{(\tau_{4})^{*}\tilde{\xi}_{2}} \\
	(\theta, \varphi, R_{\varphi} \cdot p) \ar[r]^{d\tau_{4} \qquad\quad} & (-\theta + \pi, \varphi + \pi, j_{1} \cdot R_{\varphi} \cdot p)
}\]
so that $\widetilde{d\tau_{4}}^{2}$ becomes respectively:
\begin{itemize}
	\item $e_{1}^{2}$ for $\xi_{0}$;
	\item $\tilde{R}_{-\theta} \cdot e_{1} \cdot \tilde{R}_{-\theta+\pi} \cdot \tilde{R}_{\theta-\pi} \cdot e_{1} \cdot \tilde{R}_{\theta} = e_{1}^{2}$ for $\xi_{1}$;
	\item $\tilde{R}_{-\varphi-2\pi} \cdot e_{1} \cdot \tilde{R}_{\varphi+\pi} \cdot \tilde{R}_{\theta-\pi} \cdot e_{1} \cdot \tilde{R}_{\theta} = -e_{1}^{2}$ for $\xi_{2}$.
\end{itemize}
The structures $\xi_{1}$ and $\xi_{2}$ have opposite behaviour with respect to the involution previously considered, since in this case the varible changing sign is $x$ and not $y$.

We now analyze the situation for $\tau_{3}$. First of all we can show that all the four structures satisfy $\xi_{i}\vert_{\Cyl} \simeq (\tau_{3})_{*}(\xi_{i}\vert_{T^{2} \setminus \Int(\Cyl)})$ exactly in the same way as for $\tau_{4}$, and in this case we do not have to require that the isomorphism squares to $1$. Actually, we will now prove that $\xi_{0}$ and $\xi_{1}$ (and similarly $\xi_{2}$ and $\xi_{3}$) restrict to equivalent structures on $\Cyl$, but they differ by the isomorphism $\theta$ at the boundary. In fact, the equivalence:
\begin{equation}\label{IsoCyl}
\xymatrix{
	(\theta, \varphi, p') \ar[rr]^{\rho} \ar[dr]_{\xi_{0}} & & (\theta, \varphi, \tilde{R}_{-\theta}p') \ar[dl]^{\xi_{1}} \\
	& (\theta, \varphi, p)
}
\end{equation}
is not well defined on $T^{2}$ since $\tilde{R}_{\theta} = -\tilde{R}_{\theta+2\pi}$, but if we restrict $\theta$ to the interval $[0, \pi]$, corresponding to the cylinder, there is no ambiguity left. This reasoning does not work between $\xi_{0}$ and $\xi_{2}$ since the interval of $\varphi$ is not halved. For $\xi_{0}$ we have the diagram:
	\[\xymatrix{
	(\theta, \varphi, p') \ar[r]^{(\widetilde{d\tau_{3}})_{0} \quad} \ar[d]_{\tilde{\xi}_{0}} & (-\theta, \varphi, e_{1} \cdot p') \ar[d]^{\tilde{\xi}_{0}} \ar[dl]_{(\tau_{3})^{*}\tilde{\xi}_{0}} \\
	(\theta, \varphi, p) \ar[r]^{d\tau_{3} \quad} & (-\theta, \varphi, j_{1}p)
}\]
while for $\xi_{1}$:
	\[\xymatrix{
	(\theta, \varphi, p') \ar[r]^{(\widetilde{d\tau_{3}})_{1} \qquad} \ar[d]_{\tilde{\xi}_{1}} & (-\theta, \varphi, \tilde{R}_{\theta} e_{1} \tilde{R}_{\theta} p') \ar[d]^{\tilde{\xi}_{1}} \ar[dl]_{(\tau_{3})^{*}\tilde{\xi}_{1}} \\
	(\theta, \varphi, R_{\theta} p) \ar[r]^{\quad d\tau_{3} \qquad} & (-\theta, \varphi, j_{1} R_{\theta} p)
}\]
and we can show that the two couples $(\xi_{0}, (\widetilde{d\tau_{3}})_{0}\vert_{\partial X})$ and $(\xi_{1}, (\widetilde{d\tau_{3}})_{1}\vert_{\partial X})$ are not equivalent. In fact, they are equivalent to the triples $(\xi_{0}, (\tau_{3})_{*}\xi_{0}, \id)$ and $(\xi_{1}, (\tau_{3})_{*}\xi_{1}, \id)$ via the equivalences $\rho$ of diagram \eqref{IsoCyl} and $(\tau_{3})_{*}\rho$ of the following diagram:
\begin{equation}\label{IsoCylTauStar}
\xymatrix{
	(-\theta, \varphi, e_{1}p') \ar[rr]^{(\tau_{3})_{*}\rho} \ar[dr]_{(\tau_{3})_{*}\xi_{0}} & & (-\theta, \varphi, \tilde{R}_{\theta}e_{1}p') \ar[dl]^{(\tau_{3})_{*}\xi_{1}} \\
	& (\theta, \varphi, p).
}
\end{equation}
Comparing \eqref{IsoCyl} and \eqref{IsoCylTauStar} we can see that the diagram:
	\[\xymatrix{
	(\tau_{3})_{*}\xi_{0}\vert_{\partial X} \ar[rr]^{(\tau_{3})_{*}\rho\vert_{\partial X}} & & (\tau_{3})_{*}\xi_{1}\vert_{\partial X} \\
	\xi_{0}\vert_{\partial X} \ar[rr]^{\rho\vert_{\partial X}} \ar[u]^{\id} & & \xi_{1}\vert_{\partial X} \ar[u]^{\id}
}\]
does \emph{not} commute or anti-commute. In fact, for $\theta = 0$ we get $\rho(0, \varphi, q') = (\tau_{3})_{*}\rho(0, \varphi, q')$ while for $\theta = \pi$ we get $\rho(\pi, \varphi, q') = -(\tau_{3})_{*}\rho(\pi, \varphi, q')$ since $\tilde{R}_{-\pi} = -\tilde{R}_{\pi}$. The diagram would not commute either by choosing $\rho \circ \gamma$ or $(\tau_{3})_{*}\rho \circ \gamma$ or both.

Some comments about the behaviour of $\xi_{0}, \xi_{1}, (\tau_{3})_{*}\xi_{0}, (\tau_{3})_{*}\xi_{0}$ at the boundary are needed in order to avoid possible confusion. If we embed $P_{\SO}(\partial \Cyl) \subset P_{\SO}(T^{2})$ via the outward orthogonal normal unit vector, it follows that $\{(e_{1}, e_{2}\}$ the only orthonomal oriented basis\footnote{Since the boundary has dimension $1$ there is only one oriented orthonormal basis.} at a boundary point $(0, \varphi)$ with $e_{1}$ outward, while for $(\pi, \varphi)$ the only embedded basis is $\{-e_{1}, -e_{2}\}$. Thus, since the principal bundle is $P_{\SO}(T^{2})$ is the bundle of isomorphisms from the trivial bundle $T^{2} \times \mathbb{R}^{2}$ to the tangent bundle $T(T^{2})$, which is also trivial, it follows that the embedded basis for $\theta = 0$ corresponds to $(0, \varphi, \id) \in S^{1} \times S^{1} \times \SO(2)$, while the embedded basis for $\theta = \pi$ corresponds to $(\pi, \varphi, -\id) \in S^{1} \times S^{1} \times \SO(2)$. Thus, is we consider the $\Spin$-bundles $P_{\Spin}(\partial \Cyl) \subset P_{\Spin}(T^{2})$ we have that the lifts of the embedded basis are:
	\[\begin{array}{|l|l|l|} \hline
	& \theta = 0 & \theta = \pi\\ \hline
	\textnormal{$\tilde{\xi}_{0}$-lift:} & (0, \varphi, \pm 1) & (\pi, \varphi, \pm e_{1}e_{2})\\
	\textnormal{$\tilde{\xi}_{1}$-lift:} & (0, \varphi, \pm 1) & (\pi, \varphi, \pm 1)\\
	\textnormal{$(\tau_{3})^{*}\tilde{\xi}_{0}$-lift:} & (0, \varphi, \pm e_{1}) & (\pi, \varphi, \pm (e_{1})^{2}e_{2})\\
	\textnormal{$(\tau_{3})^{*}\tilde{\xi}_{1}$-lift:} & (0, \varphi, \pm e_{1}) & (\pi, \varphi, \pm e_{1})\\ \hline
\end{array}\]
It may seem strange that at the boundary, whose tangent space is generated only by $e_{2}$, also the outward vector $e_{1}$ is involved, but that's due to the fact that on $\pi$ there is a $-1$ to lift due to the orientation and for all the structures different from $\tilde{\xi}_{0}$ there is a twist in the projection of the third factor $\Spin(2) \rightarrow \SO(2)$ which makes $e_{1}$ enter in the lifting. The isomorphisms of spin structures we dealt with until now are then at the boundary:
	\[\begin{array}{lll}
	\rho(0, \varphi, \pm 1) = (0, \varphi, \pm 1) & & \rho(\pi, \varphi, \pm e_{1}e_{2}) = (\pi, \varphi, \mp 1)\\
	(\widetilde{d\tau_{3}})_{0}(0, \varphi, \pm 1) = (0, \varphi, \pm e_{1}) & & (\widetilde{d\tau_{3}})_{0}(\pi, \varphi, \pm e_{1}e_{2}) = (\pi, \varphi, \pm (e_{1})^{2}e_{2})\\
	(\widetilde{d\tau_{3}})_{1}(0, \varphi, \pm 1) = (0, \varphi, \pm e_{1}) & & (\widetilde{d\tau_{3}})_{1}(\pi, \varphi, \pm 1) = (\pi, \varphi, \pm e_{1}) \; .
\end{array}\]
We know that $\tilde{\xi}_{0} \cup_{\id} (\tau_{3})^{*}\tilde{\xi}_{0} \simeq \tilde{\xi}_{0} \cup_{(\widetilde{d\tau_{3}})_{0}} \tilde{\xi}_{0}$ and the same for $\tilde{\xi}_{1}$. Some representatives of the two equivalence classes are then:
	\[\begin{array}{ll}
	\textnormal{Class inducing $\tilde{\xi}_{0}$ on $T^{2}$:} & \tilde{\xi}_{0} \cup_{(\widetilde{d\tau_{3}})_{0}} \tilde{\xi}_{0} \; \simeq \; \tilde{\xi}_{0} \cup_{\id} (\tau_{3})^{*}\tilde{\xi}_{0} \; \simeq \; \tilde{\xi}_{1} \cup_{\id} \tilde{\xi}_{1} \\
	\textnormal{Class inducing $\tilde{\xi}_{1}$ on $T^{2}$:} & \tilde{\xi}_{1} \cup_{(\widetilde{d\tau_{3}})_{1}} \tilde{\xi}_{1} \; \simeq \; \tilde{\xi}_{1} \cup_{\id} (\tau_{3})^{*}\tilde{\xi}_{1} \; \simeq \; \tilde{\xi}_{0} \cup_{\id} \tilde{\xi}_{0} \; .
\end{array}\]
It is easy to find the invariance conditions for pinors and spinors as sections of the associated vector bundles. Moreover, all this picture is equivalent to considering $(T^{2}, \tau')$; we leave the details to the reader.

\section*{Acknowledgements}

We would like to thank Ludwik Dabrowski for useful discussions.


\end{document}